\documentclass[aps,prl,twocolumn,english,superscriptaddress,longbibliography,reprint]{revtex4-2}
\usepackage[mathlines]{lineno}

\usepackage[utf8]{inputenc}
\usepackage{graphicx}
\usepackage{times}
\usepackage{xcolor}
\usepackage{tikz}
\usetikzlibrary{matrix,calc}
\usepackage{amsthm}
\usepackage{amsmath}
\usepackage{amssymb}
\usepackage{bm}
\usepackage{enumitem}
\usepackage{mathrsfs}
\usepackage{verbatim}
\usepackage{braket}
\usepackage{lipsum}
\usepackage{url}
\usepackage{tikz}
\usetikzlibrary{matrix,calc}
\usepackage{physics}
\usepackage{swphysics}
\usepackage{mathtools}
\usepackage{float}

\usepackage{hyperref}

\definecolor{mycitecolor}{HTML}{4169E1}   
\definecolor{mylinkcolor}{HTML}{9C27B0}   

\definecolor{gateblue}{RGB}{143,213,236}
\definecolor{gatepurple}{RGB}{207,158,235}
\definecolor{gateborder}{RGB}{105,105,105}
\hypersetup{
    colorlinks=true,
    citecolor=mycitecolor,
    linkcolor=mylinkcolor,
    urlcolor=blue
}

\NewDocumentCommand{\pic}{ O{scale=1} O{0em} m }{%
    \vcenter{\hbox{\raisebox{#2}{\includegraphics[#1]{figures/#3.pdf}}}}%
}

\NewDocumentCommand{\picR}{ O{scale=1} O{0em} m }{%
  \mathord{%
    \raisebox{#2}{%
      \ensuremath{%
        \vcenter{\hbox{%
          \includegraphics[#1]{figures/#3.pdf}%
        }}%
      }%
    }%
  }%
}
\newsavebox{\myobjbox}

\newcommand{\upd}[1]{^\mathrm{#1}}
\newcommand{\ind}[1]{_\mathrm{#1}}
\DeclareMathOperator{\sgn}{\mathrm{sgn}}
\newcommand{\todoF}[1]{\textbf{\color{blue} [F: #1]}}

\newcommand{\todoU}[1]{\textbf{\color{red} [U: #1]}}

\newcommand{\sw}[1]{\textcolor{orange}{#1}}

\makeatletter
\newcommand{\SMcontents}{%
  \begin{center}\textbf{CONTENTS}\end{center}%
  \@starttoc{smt}%
}
\makeatother

\newcommand{\SMsection}[1]{%
  \section{#1}%
  \addcontentsline{smt}{section}{%
    \protect\numberline{\thesection}\hspace{0.5em}#1}%
}
\begin{document}

\title{Kardar--Parisi--Zhang superdiffusion in chaotic quantum circuits}
\author{Rustem Sharipov}
\email{rustem.sharipov@fmf.uni-lj.si}
\affiliation{Department of Physics, Faculty of Mathematics and Physics, University of Ljubljana, 1000 Ljubljana, Slovenia}
\author{Urban Duh}
\email{urban.duh@fmf.uni-lj.si}
\affiliation{Department of Physics, Faculty of Mathematics and Physics, University of Ljubljana, 1000 Ljubljana, Slovenia}
\author{Sun Woo P. Kim}
\thanks{ \texttt{swk34} \texttt{[at]} \texttt{cantab} \texttt{[dot]} \texttt{ac} \texttt{[dot]} \texttt{uk}}
\affiliation{Department of Physics, King's College London, Strand, London WC2R 2LS, United Kingdom}
\author{Friedrich Hübner}
\email{friedrich.huebner@phys.ens.fr}
\affiliation{Laboratoire de Physique de l’École Normale Superieure, CNRS, ENS \& Université PSL, Sorbonne Université, Université Paris Cité, 75005 Paris, France}

\begin{abstract}
We construct a family of chaotic brickwork quantum circuits exhibiting
superdiffusive transport with dynamical exponent $z=3/2$ in the full
Kardar--Parisi--Zhang (KPZ) universality class. This is enabled by the
equilibrium current associated with a inhomogeneous conserved density having a
nontrivial dependence on the chemical potential. Using nonlinear fluctuating
hydrodynamics, we derive the propagation velocity, broadening scale, and
universal scaling form of the charge correlation function from microscopic
equilibrium properties. Tensor-network simulations of a two-site spin-$1$ (qutrit) circuit
reproduce these parameter-free predictions, including the full stationary KPZ
profile. We further construct a three-site spin-$\frac12$ (qubit) circuit, where transport varies
from diffusive to superdiffusive with the chemical potential, while long-lived
coherent quasiparticles delay the convergence of the profile to the KPZ
scaling form. Our results establish that KPZ superdiffusion in quantum systems is not restricted to fine-tuned integrable models.
\end{abstract}

\maketitle

\textit{\textbf{Introduction}---} 
Conservation laws govern slow relaxation processes in ergodic many-body systems. 
The long-time and long-wavelength dynamics towards equilibrium is described by hydrodynamics, which predicts the spatiotemporal correlations of the conserved densities, characterised by a typical length scale growing with the dynamical exponent $z$ as $\sim t^{1/z}$. In generic chaotic systems, these correlations are expected to form ballistic propagating peaks ($z=1$) that broaden diffusively ($z=2$). Transport deviating from this standard expectation is referred to as anomalous \cite{RevTransport}. Several mechanisms of anomalous transport have been identified, including kinetic constraints \cite{Subdiffusion_kinetic,PXP_Super}, multipole conservation \cite{Subdiffusion_Dipole,Subdiff_anyd}, and nodal points \cite{Nodal_orig,Nodal_chaos}. But perhaps the best-known theory that explains a broad class of anomalous transport in one-dimensional systems is nonlinear fluctuating hydrodynamics (NLFHD). It predicts superdiffusive dynamical exponents when equilibrium currents depend nonlinearly on the thermodynamic fields. The leading nontrivial case $z=3/2$ belongs to the Kardar--Parisi--Zhang (KPZ) universality class (first studied in context of surface growth \cite{KPZ_orig,KPZintroduction}) appearing in classical fluids and anharmonic chains \cite{MendlSpohn2013,spohn2014,Bertini1997} and recently cellular automata \cite{Kim2026,Sharipov2026}.
\begin{figure}[h!]
    \centering
\includegraphics[width=0.5\textwidth]{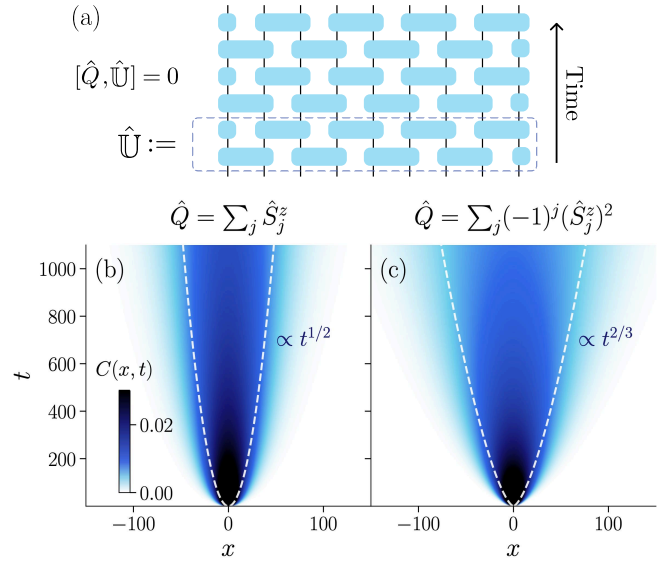}
\caption{\textbf{Diffusive and superdiffisive spreading of a conserved charge.}
(a) Setup: Time-evolution $\hat{\mathbb{U}}$ consists of two layers of two-qutrit gates arranged in a brick-wall geometry. The circuit conserves a single conserved quantity satisfying $[\hat{Q},\hat{\mathbb{U}}]=0$.
(b) For a homogeneous charge,
the charge density correlator $C(x,t)$ spreads diffusively as
$\sim t^{1/2}$.
(c) For a staggered charge, $C(x, t)$ exhibits KPZ
superdiffusion $\sim t^{2/3}$.
The white dashed curves show the location of the half-maximum of $C(x, t)$.}
    \label{fig:main}
\end{figure}

Despite numerous classical examples of KPZ superdiffusion, 
generic quantum spin chains are diffusive. 
Integrable systems 
provide an important exception \cite{ljubotina2017spin}: the Heisenberg XXX chain exhibits superdiffusion with $z=3/2$, and its correlation profile closely resembles the KPZ scaling function \cite{Ljubotina2019}. Yet, its full counting statistics are inconsistent with standard KPZ theory \cite{Ziga_non_KPZ}, as also observed in Google's quantum-processor experiment \cite{rosenberg2024dynamics}. Furthermore, the superdiffusive behaviour is restricted to fine-tuned integrable points and originates from large quasiparticle bound states associated with non-Abelian symmetry \cite{Enej_Super}, a fundamentally different mechanism to that of NLFHD. In fact, it is well known~\cite{PhysRevLett.129.176601,Watanabe2019}, and discussed in the Supplementary Material (SM)~[\onlinecite{SuppMat}], that Hamiltonian chains with a single on-site conserved charge have identically vanishing equilibrium currents, precluding the NLFHD mechanism for superdiffusion. Currently, there are no known examples of non-integrable quantum systems which display superdiffusive transport in the full KPZ universality, which raises the question of whether they exist at all.

In this letter, we demonstrate that it is indeed possible to obtain KPZ
superdiffusive transport in chaotic quantum spin chains if the time evolution is
not continuous, but instead discrete. We construct a family of chaotic brickwork
quantum circuits with a single conserved charge, where each gate of the circuit
acts on $r=2$ or $3$ sites, see Fig.~\ref{fig:main}(a). Brickwork quantum
circuits are natural ways to simulate quantum spin chains on quantum
computers, and hence our proposed models can be straightforwardly
implemented on various quantum platforms~\cite{rosenberg2024dynamics,google2025observation,Hoke2023,Koh2023,PhysRevResearch.4.023097,fischer2026dynamical}. In the constructed circuits, the conserved density shares the same spatial
unit cell of the Floquet dynamics: both total charge $\hat Q$ and the Floquet evolution $\hat{\mathbb{U}}$ are invariant under translation by $r$ sites, not generally under a one-site
translation. This structure permits a nonlinear
equilibrium current for generic circuits. Within our construction, all $r=2$
realizations for spin-$\tfrac12$ (qubits) are either integrable or have zero equilibrium current~[SM \onlinecite{SuppMat}, \onlinecite{Urban_integr}]; the minimal chaotic models therefore employ either $r=2$ gates for
spin-$1$ (qutrits) systems or $r=3$ gates for spin-$\tfrac12$ ones. Starting from
the microscopic charge density and current operators, by employing NLFHD, we analytically
calculate the equilibrium current, susceptibility, propagation velocity, and
nonlinear coupling that determine the complete stationary KPZ correlation
function. In the spin-1 case, tensor-network simulations converge to these
predictions without any fitting to amplitudes, velocities, or spatial scales. In
the spin-$\frac{1}{2}$ case, we predict KPZ superdiffusion only at nonzero chemical
potential. Its emergence at large $\mu$ is slowed down due to a crossover with
``low-energy'' quantum excitations we explicitly identify.

\begin{figure}[t]
    \centering

\includegraphics[width=0.48\textwidth]{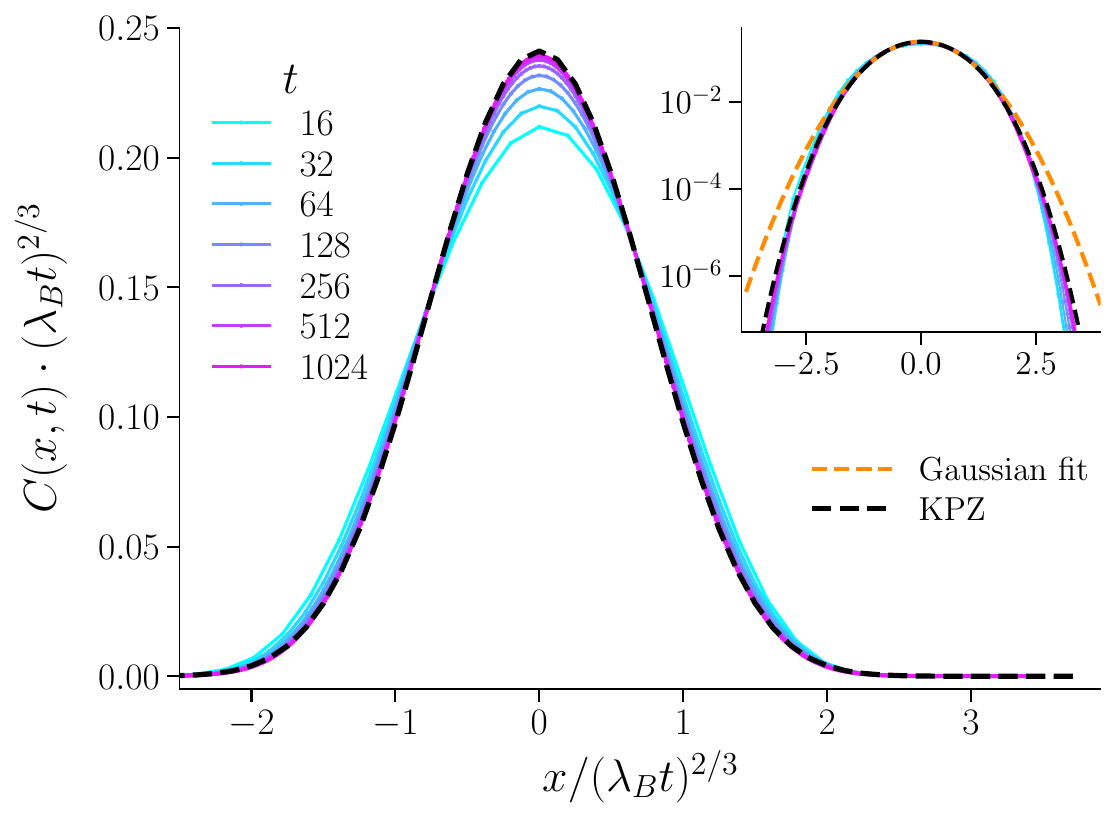}
    \caption{\textbf{KPZ scaling collapse for the staggered conserved charge in the chaotic spin-$\boldsymbol 1$  circuit}. TEBD results for the connected correlation function $C(x,t)=\langle \hat q(x,t)\hat q(0,0)\rangle^\mathrm   {c}$ are plotted as $(\lambda_B t)^{2/3}C(x,t)$ versus $x/(\lambda_B t)^{2/3}$, with color indicating the evolution time. As $t$ increases, the numerical profiles collapse onto the exact stationary KPZ scaling function shown in black dashed line. The normalization $c_0=4/9$ and superdiffusion scale $\lambda_B=1/\sqrt{2}$ are determined analytically. The bond dimension of TEBD simulation $\chi=256$. The inset shows the same data on a semilogarithmic scale and orange dashed line indicates best Gaussian fit of the profile.}
    \label{fig:Corr_profile}
\end{figure}

\textit{\textbf{Model and main results}---}The simplest example of a circuit with superdiffusive transport is given by the following family of models. We consider a chain of spin-$1$ degrees of freedom with local states $\ket{\uparrow},\ket{\downarrow} ~\text{and} ~\ket{0}$ evolving under the homogeneous brickwork circuit illustrated in Fig.~\ref{fig:main}(a), such that each gate acts on $r=2$ sites. The time evolution is given by the repetition of two layers of evolution, given by 
\begin{equation} 
    \hat{\mathbb U} := \hat{\mathbb U}_{\mathrm{o}}\hat{\mathbb U}_{\mathrm{e}}, \label{eq:floquet} 
\end{equation} 
where
\begin{equation} 
    \hat{\mathbb U}_{\mathrm{o}} := \prod_{n} \hat U_{2n+1,2n+2}, \qquad \hat{\mathbb U}_{\mathrm{e}} := \prod_{n} \hat U_{2n,2n+1},
\end{equation}
with indices denoting on which sites the operators act.
The gate $\hat{U}$ is given by $\hat{U}=\hat{V}\hat{S}$, where
\begin{equation}
\hat V =
\resizebox{0.65\columnwidth}{!}{%
\begin{tikzpicture}[
    baseline=(M.center),
    blockline/.style={
        draw=black!35,
        dashed,
        line width=0.35pt,
        dash pattern=on 2pt off 1.3pt
    }
]
\matrix (M) [
    matrix of math nodes,
    ampersand replacement=\&,
    left delimiter={(},
    right delimiter={)},
    nodes={
        anchor=center,
        inner xsep=1.7pt,
        inner ysep=1.7pt
    },
    column sep=0pt,
    row sep=0pt
] {
    {}
    \& \bra{0\small{\uparrow}}
    \& \bra{0\small{\downarrow}}
    \& \bra{\small{\uparrow} 0}
    \& \bra{\small{\downarrow} 0}
    \& \bra{00}
    \& \bra{\downarrow\downarrow}
    \& \bra{\downarrow\uparrow}
    \& \bra{\uparrow\downarrow}
    \& \bra{\uparrow\uparrow} \\
    \ket{0\small{\uparrow}}
    \& \ast \& \ast \& {} \& {} \& {} \& {} \& {} \& {} \& {} \\
    \ket{0\small{\downarrow}}
    \& \ast \& \ast \& {} \& {} \& {} \& {} \& {} \& {} \& {} \\
    \ket{\small{\uparrow} 0}
    \& {} \& {} \& \ast \& \ast \& {} \& {} \& {} \& {} \& {} \\
    \ket{\small{\downarrow} 0}
    \& {} \& {} \& \ast \& \ast \& {} \& {} \& {} \& {} \& {} \\
    \ket{00}
    \& {} \& {} \& {} \& {} \& \ast \& \ast \& \ast \& \ast \& \ast \\
    \ket{\downarrow\downarrow}
    \& {} \& {} \& {} \& {} \& \ast \& \ast \& \ast \& \ast \& \ast \\
    \ket{\downarrow\uparrow}
    \& {} \& {} \& {} \& {} \& \ast \& \ast \& \ast \& \ast \& \ast \\
    \ket{\uparrow\downarrow}
    \& {} \& {} \& {} \& {} \& \ast \& \ast \& \ast \& \ast \& \ast \\
    \ket{\uparrow\uparrow}
    \& {} \& {} \& {} \& {} \& \ast \& \ast \& \ast \& \ast \& \ast \\
};

\coordinate (xleft) at (M-1-2.west);

\coordinate (xmiddleone) at
    ($(M-1-3.east)!0.5!(M-1-4.west)$);

\coordinate (xmiddletwo) at
    ($(M-1-5.east)!0.5!(M-1-6.west)$);

\coordinate (xright) at (M-1-10.east);

\coordinate (ytop) at
    ($(M-1-2.south)!0.5!(M-2-2.north)$);

\coordinate (ymiddleone) at
    ($(M-3-2.south)!0.5!(M-4-2.north)$);

\coordinate (ymiddletwo) at
    ($(M-5-2.south)!0.5!(M-6-2.north)$);

\coordinate (ybottom) at (M-10-1.south);

\draw[blockline]
    (xleft |- ytop)
    --
    (xleft |- ybottom);

\draw[blockline]
    (xmiddleone |- ytop)
    --
    (xmiddleone |- ybottom);

\draw[blockline]
    (xmiddletwo |- ytop)
    --
    (xmiddletwo |- ybottom);
    
\draw[blockline]
    (xright |- ytop)
    --
    (xright |- ybottom);

\draw[blockline]
    (xleft |- ytop)
    --
    (xright |- ytop);

\draw[blockline]
    (xleft |- ymiddleone)
    --
    (xright |- ymiddleone);

\draw[blockline]
    (xleft |- ymiddletwo)
    --
    (xright |- ymiddletwo);
    
\draw[blockline]
    (xleft |- ybottom)
    --
    (xright |- ybottom);
\end{tikzpicture}%
}
\label{eq:block_gate}
\end{equation}
and $\hat{S} = \sum_{a,b=\uparrow,\downarrow,0} \ket{b}\ket{a}\bra{a}\bra{b}$ is the two-site swap gate. All of the blocks indicated by $*$ can be arbitrary unitary matrices, as long as they are not near points that host further conserved charges. The concrete model used to generate Fig.~\ref{fig:Corr_profile} is given in the SM~\cite{SuppMat}.

As we will derive in the next section, this circuit conserves the following `staggered' conserved quantity,
\begin{align}
    \hat{Q} &= \sum_n \hat{q}\upd{o}_{2n-1} + \hat{q}\upd{e}_{2n} = \sum_n (-1)^n (\hat{S}^z_n)^2, \label{eq:charge_stag}
\end{align}
where $S^z = \ketbra{\uparrow} -\ketbra{\downarrow}$.
Note that the densities $\hat q^{e/o}$ on even and odd sites differ.

As it will become clear in the next section, the staggering of the charge is essential for superdiffusion.
Based on our theoretical analysis, we predict that such circuits show infinite-temperature ($\hat{\rho} \sim \hat{\mathbb{I}}$) charge-density correlators with superdiffusive dynamical exponent $z=3/2$ at asymptotically large evolution times and spatial scales:
\begin{align}
C(x,t) := \expval{\hat{q}(x,t)\hat{q}(0,0)}\upd{c} =  \frac{c_0}{(\lambda\ind{B}t)^{2/3}} f\ind{KPZ}\qty(\frac{x}{(\lambda\ind{B}t)^{2/3}}), \label{eq:corr_KPZ}
\end{align}
where  $\hat{q}(x): = \hat{q}\upd{o}_{2x-1} + \hat{q}\upd{e}_{2x}$ is the charge density, $\hat{q}(x,t)=(\hat{\mathbb{U}}^{\dagger})^t \hat{q}(x) \hat{\mathbb{U}}^t$ and $\langle \hat{A}\hat{B}\rangle^\mathrm{c} = \langle \hat{A}\hat{B}\rangle-\langle \hat{A} \rangle \langle \hat{B} \rangle$. The normalization constant $c_0=4/9$ and the superdiffusion constant $\lambda\ind{B} = 1/\sqrt{2}$ are explicitly computed in the End Matter and the KPZ shape $f\ind{KPZ}\qty(x)$ is known from the exact solution of the KPZ equation~\cite{Sasamoto_Spohn}.

The asymptotic form \eqref{eq:corr_KPZ} is confirmed in Fig.~\ref{fig:Corr_profile} against numerical TEBD simulations [SM~\onlinecite{SuppMat}] which shows excellent agreement.

\textit{\textbf{Conservation laws for many-body circuits}---}
We now derive a condition for a circuit to conserve the charge of the form \eqref{eq:charge_stag}. First, observe that we can write $\hat{\mathbb{U}} = (\hat{\mathbb{U}}_o\hat{T})^2\hat{T}^{-2}$, where $\hat{T}$ is the right shift operator. Since $\hat{Q}$ commutes with $\hat{T}^2$, if $\hat{Q}$ commutes with $\hat{\mathbb{U}}_o\hat{T}$ it will also commute with $\hat{\mathbb{U}}$. Thus, we require
\begin{align}
    \hat{\mathbb{U}}_o^\dagger \hat{Q}\hat{\mathbb{U}}_o &= \hat{T}\hat{Q}\hat{T}^\dagger \label{eq:q_drift}
\end{align}
represented pictorially as
\begin{align}
\begin{aligned}
& \sum_{i \mathrm{\; odd}} \cdts~ \picR[scale=0.024][-0.6em]{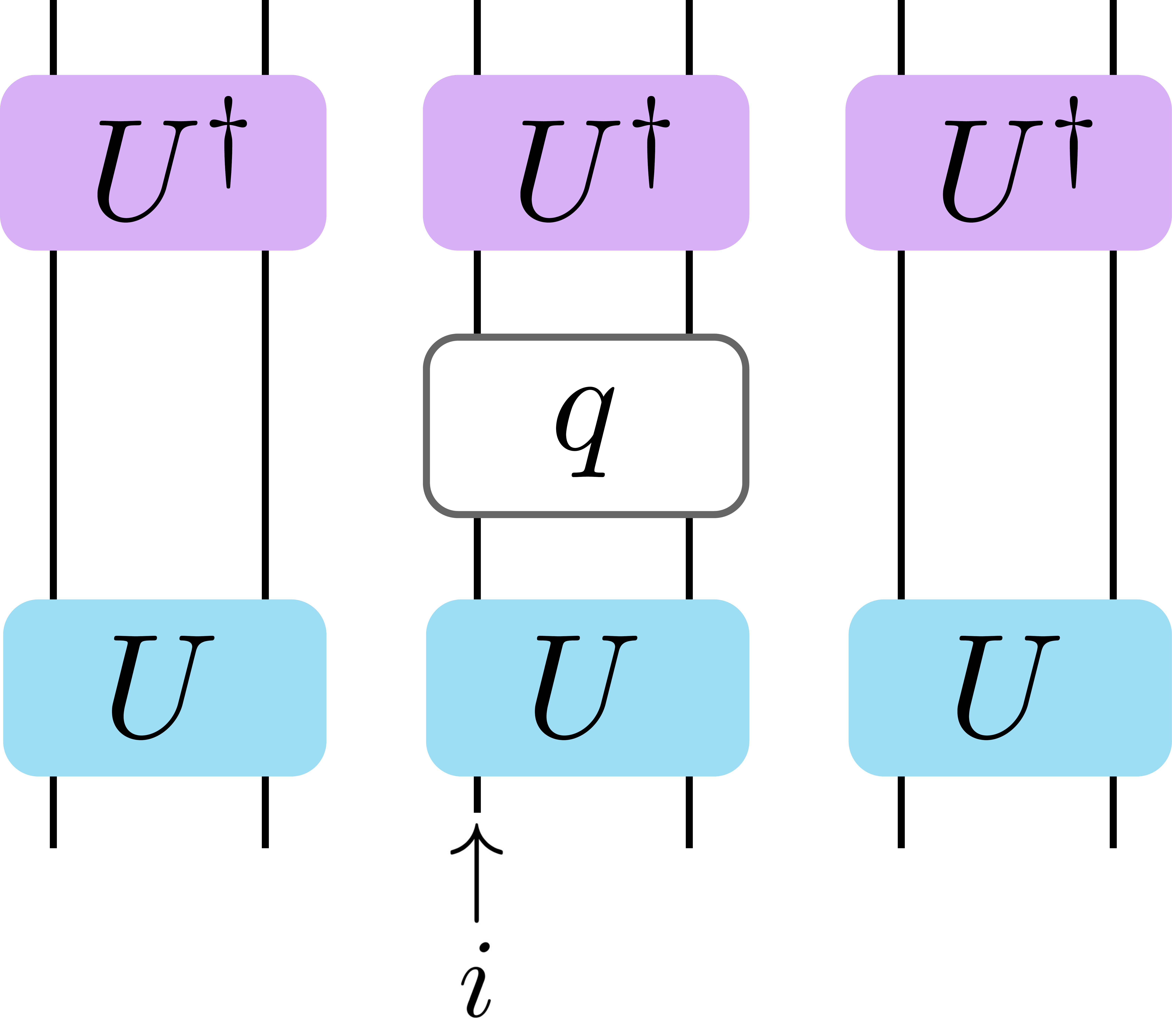} ~\cdts = \sum_{i \mathrm{\; odd}} \cdts ~\, \picR[scale=0.024][-0.8em]{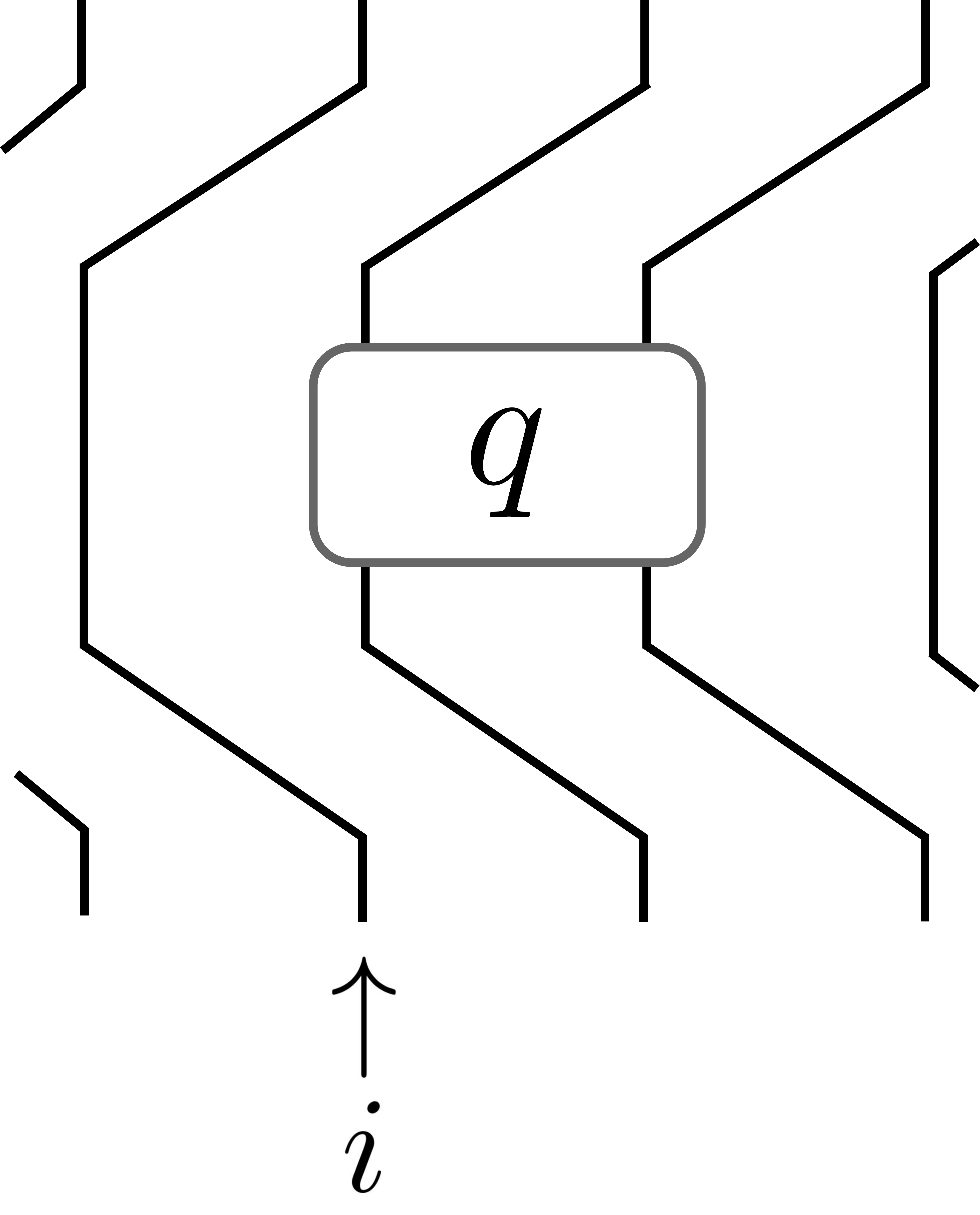} \, ~\cdts \,.\nonumber
\end{aligned}
\end{align}
The above equation implies
\begin{align}
\begin{aligned}
& \picR[scale=0.024][-0.1em]{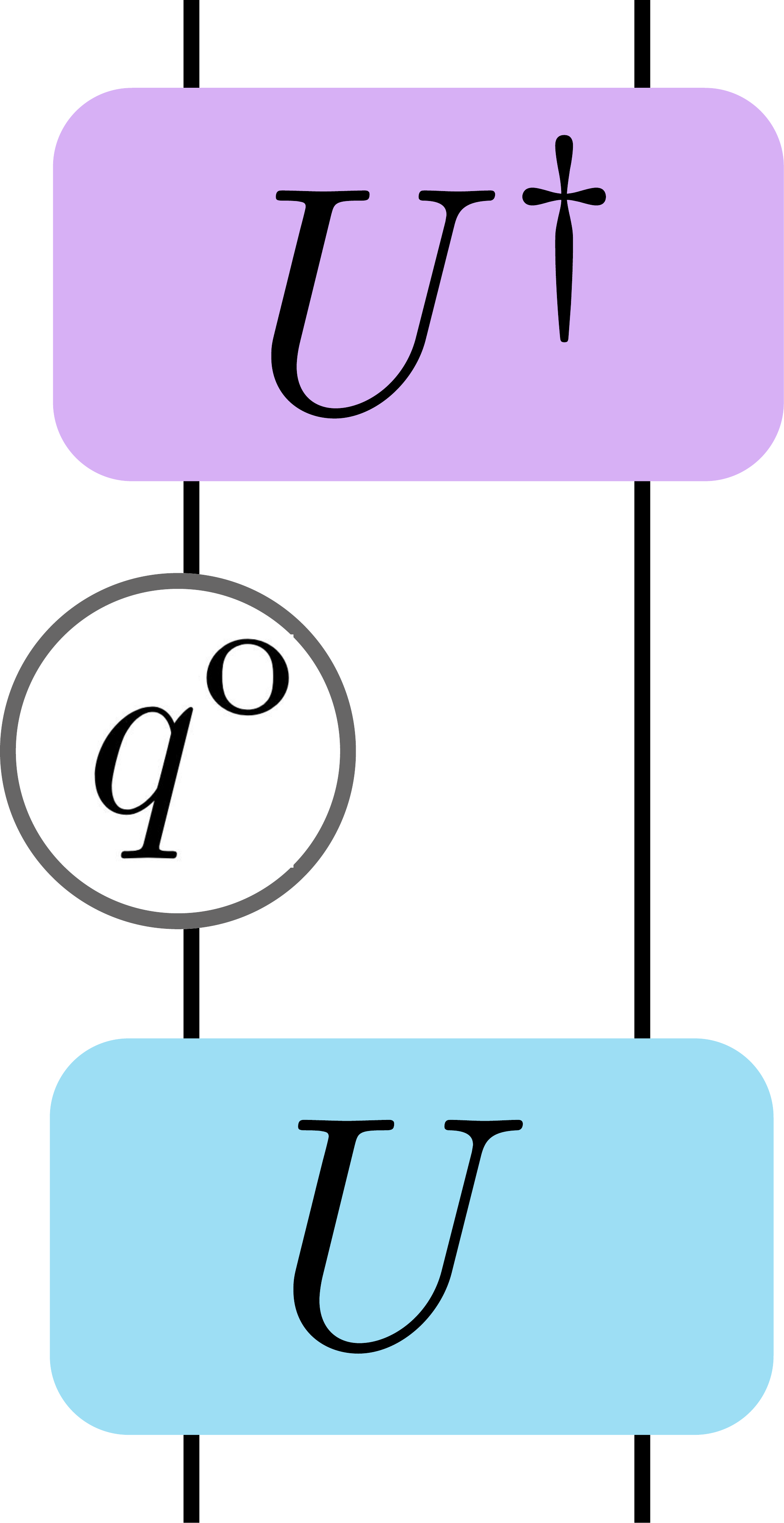}~~+  ~~\picR[scale=0.024][-0.1em]{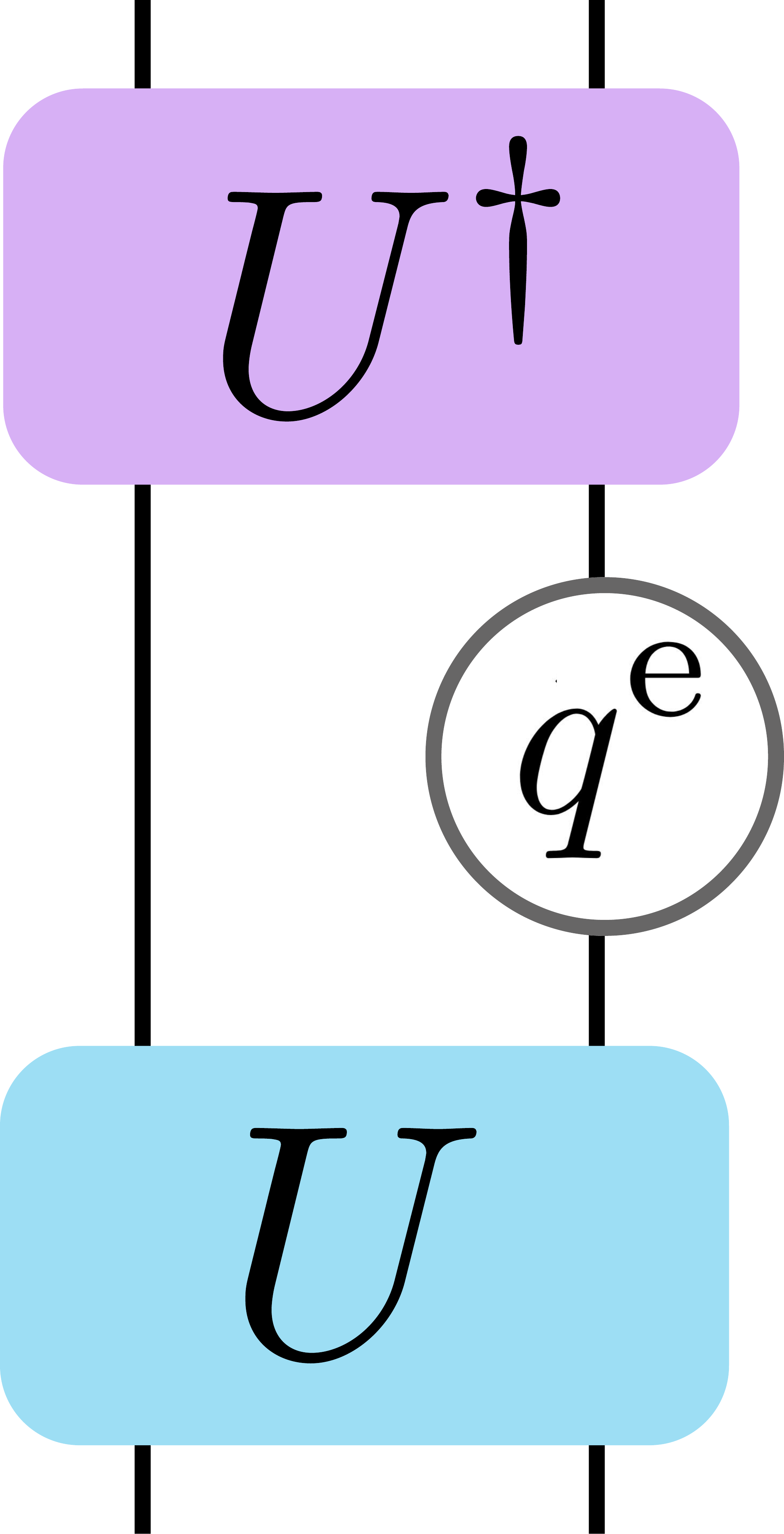}~=~~ \, \picR[scale=0.024][-0.1em]{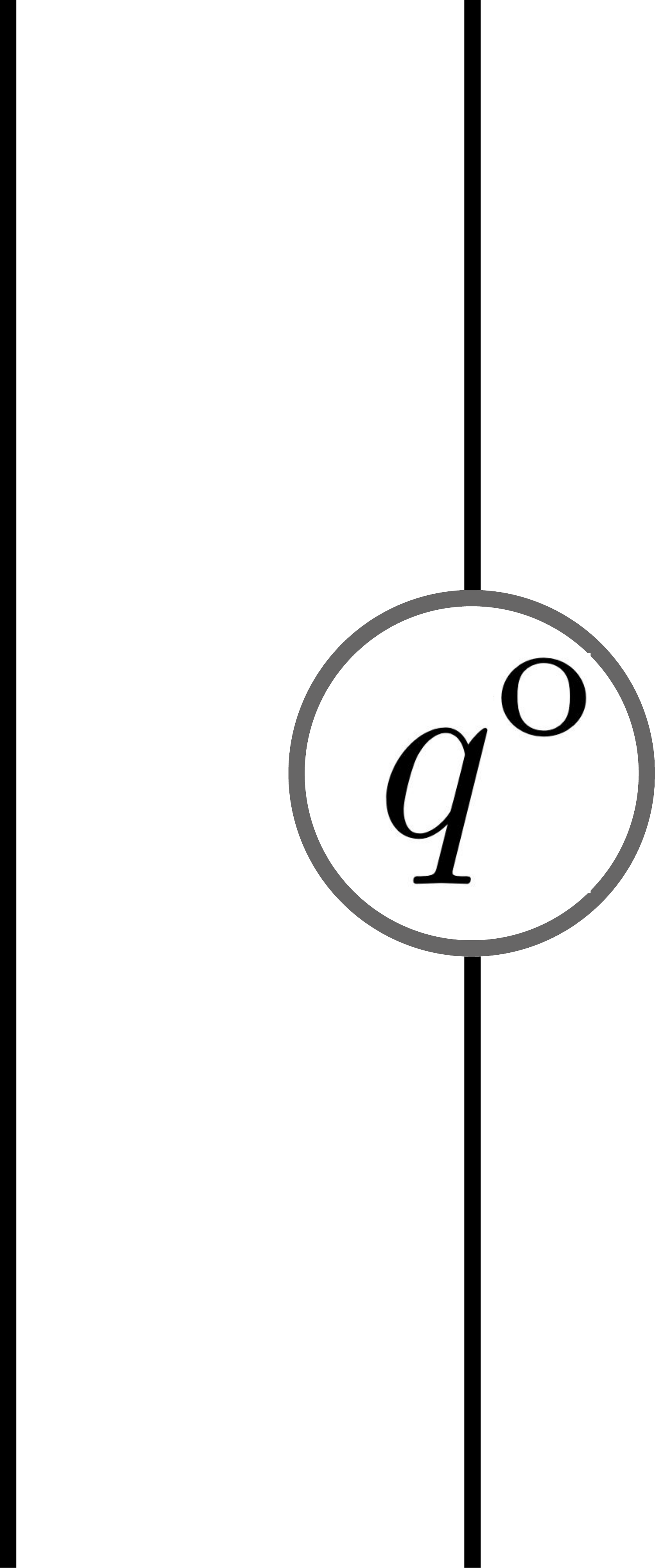}~~+~~\picR[scale=0.024][-0.1em]{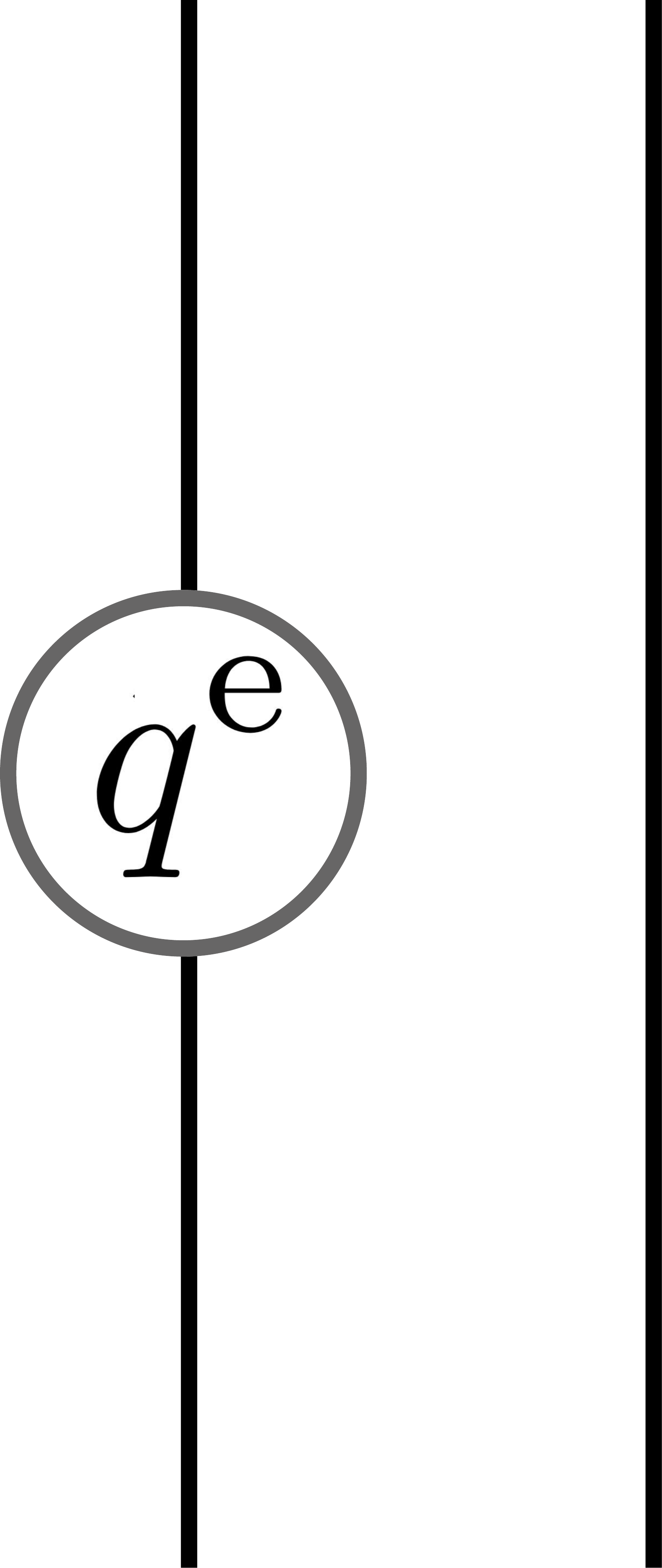}~~,
\end{aligned}
\end{align}
which can be rewritten as
    \begin{align}\label{VqVq}
\begin{aligned}
& \picR[scale=0.024][-0.0em]{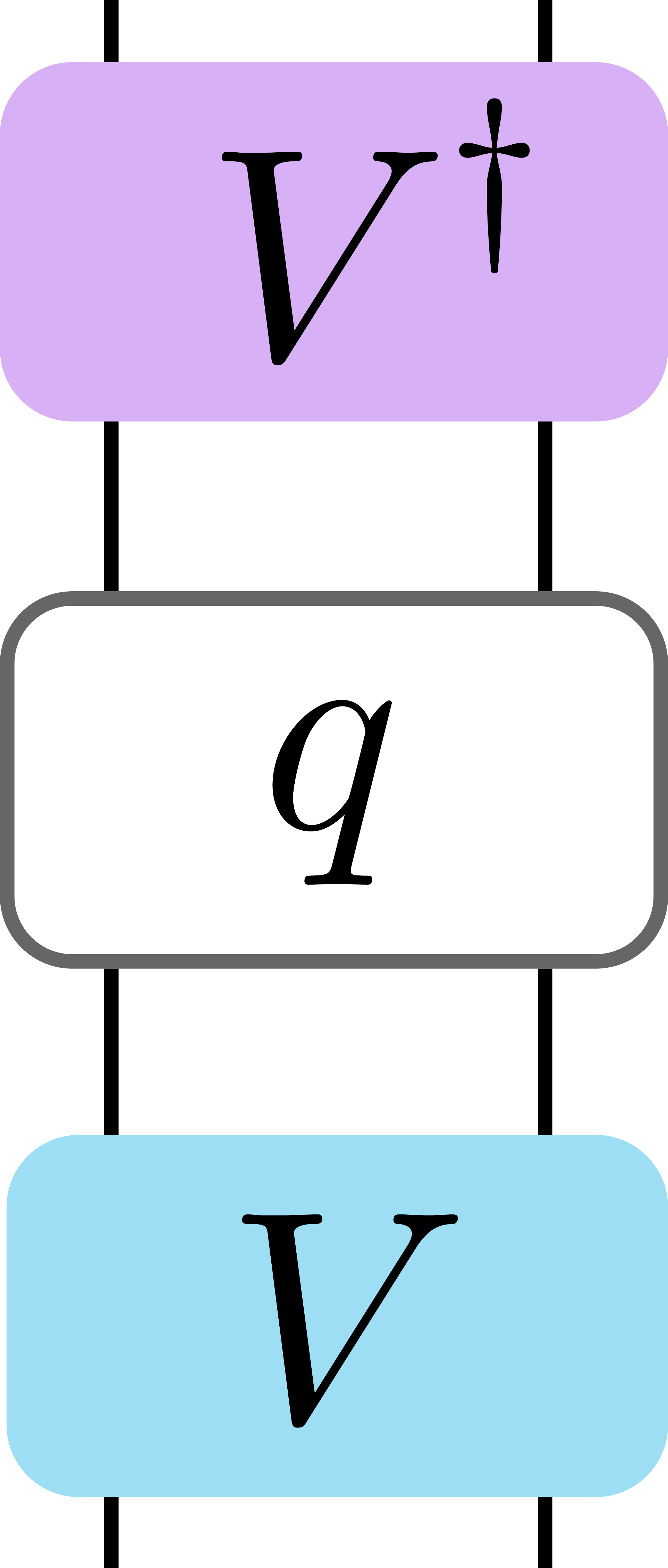}~~=~~ \, \picR[scale=0.025][-0.0em]{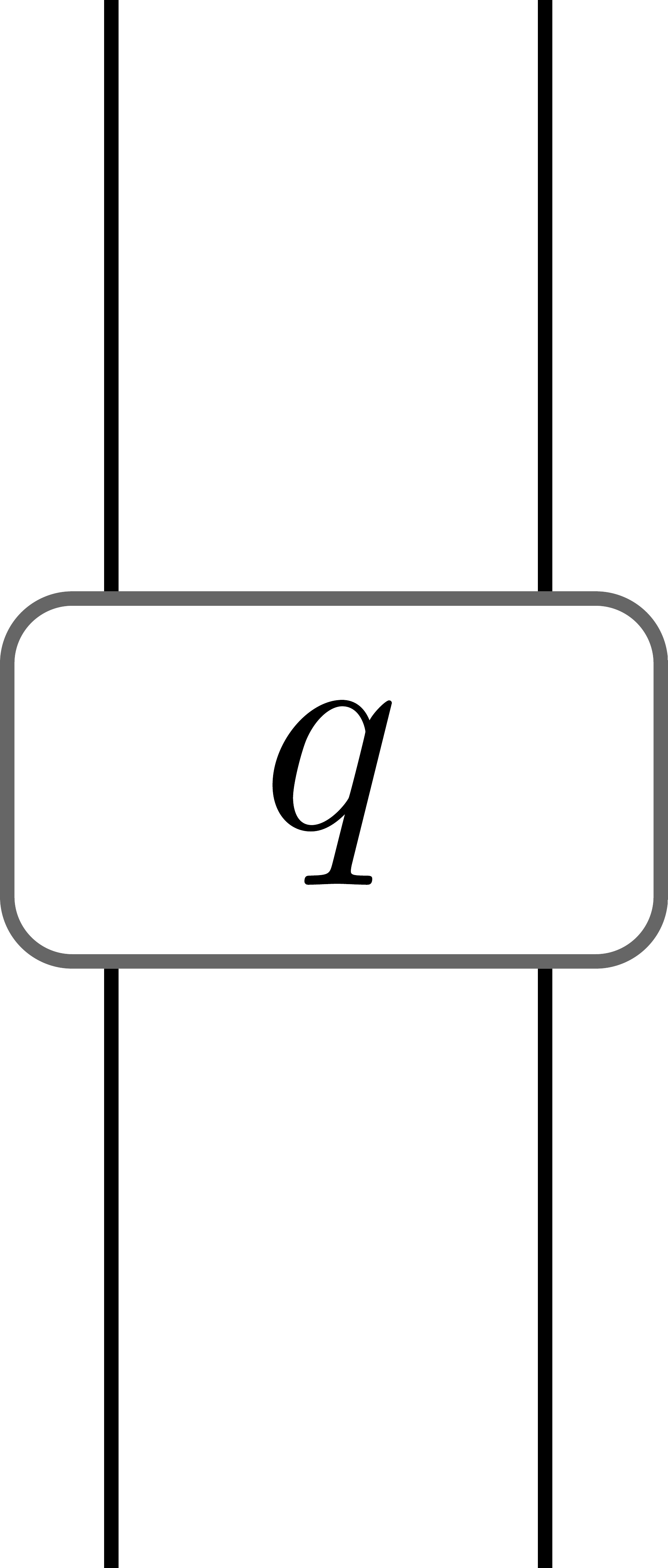}~,~~\text{where} & \picR[scale=0.025][-0.1em]{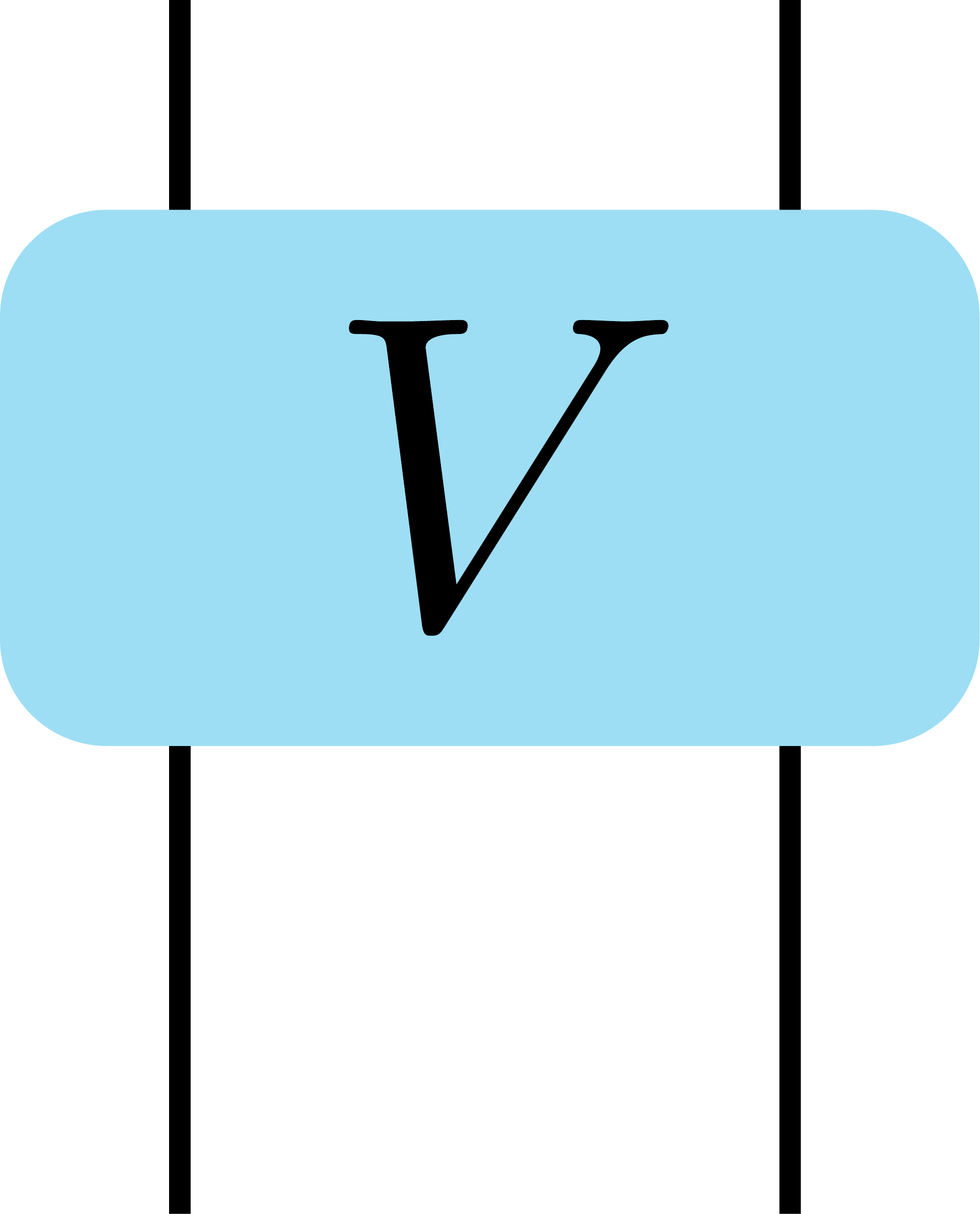}~~:=~~\picR[scale=0.024][-0.3em]{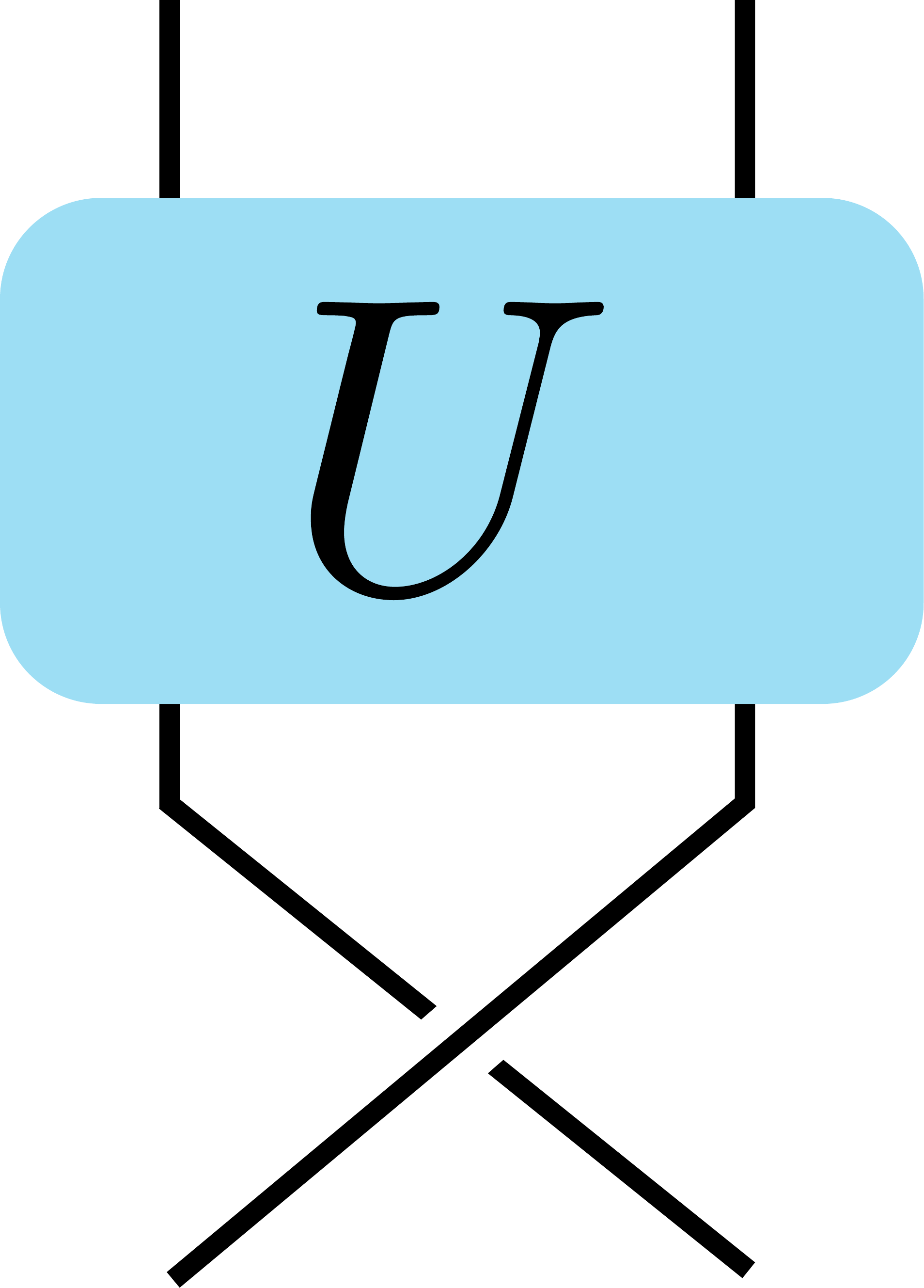}~
\end{aligned}
\end{align}
or in formula, $\hat{V} = \hat{U}\hat{S}$. Eq.~\eqref{VqVq} is equivalent to $[\hat{V},\hat{q}] = 0$, meaning that $\hat{V}$ is block-diagonal in the charge sectors of $\hat{q}$ shown in Eq.~\eqref{eq:block_gate}. Thus, given the charge $\hat{Q}$ of the form \eqref{eq:charge_stag}, we can easily construct gates that conserves this charge. Note that some gates might have further charges, e.g., for $\hat{V}=\hat{I}$ we obtain the super-integrable SWAP circuit. Such fine-tuned circuits will then exhibit a different behaviour of correlation functions.

As a second comment, note that unless the charge $\hat{Q}$ is invariant under one-site translations, this circuit construction has no continuous-time limit: the finite SWAP factor in $\hat U=\hat V\hat S$ prevents writing $\hat U=e^{-i\hat h\delta t}$ with a finite local generator as $\delta t\to0$. It is thus more natural to think of the gate $\hat{U}$ as a deformation of the SWAP gate breaking all its charges except $\hat{Q}$.

\textit{\textbf{Coarse-grained description}---} To explain the origin of superdiffusion, we follow the ideas of nonlinear fluctuating hydrodynamics (NLFHD)~\cite{spohn2014}.

We want to understand the dynamics of correlation functions, which are related
to the dynamics of equilibrium states $\hat{\rho}_\mu \sim
e^{\mu\hat{Q}}$ with slight perturbations within the framework of linear response~\cite{pottier, RevTransport}. Hydrodynamics gives a coarse-grained
description of the dynamics of the system at times long enough such that the system can be described as locally but not globally, equilibrated~\cite{landau_fluid_mechanics,Pines1966,spohn2012large,doyon2020lecture,Kim2026}, i.e., for local observables, the state is well-approximated by $\hat{\rho}\sim e^{\sum_x \mu(x,t) \hat{q}(x)}$ with a slowly varying chemical potential $\mu(x,t)$.
Defining the coarse-grained charge
density $q(x, t) := \ev*{\hat q}_{\mu(x, t)}$ and the corresponding current
$\jmath(x, t) := \ev*{\hat \jmath}_{\mu(x, t)}$, where $\ev{\, \cdot \,}_{\mu} := \mathrm{tr}[\, \cdot \, \hat \rho_\mu]$, we arrive at the coarse-grained
continuity equation (derivation in the End matter)
\begin{align}
    \partial_t q(x, t) + \partial_x \jmath(x, t) &= 0.\label{eq:Euler} 
\end{align}
One can now recast $\jmath$ to depend on $q$ in the following way. For a given
$q$, find the corresponding local equilibrium state $\hat{\rho}_\mu$ by matching
the chemical potential $\mu$ such that $q=q(\mu)=\expval{\hat{q}}_\mu$,
then set $\jmath(q) = \expval{\hat{\jmath}}_\mu$. Next, we may expand over a background with constant chemical potential $q_0 := \ev{\hat
q}_{\mu}$, $q(x, t) = q_0 + \delta q(x, t)$, to obtain
\begin{equation}
    \partial_t \delta q + \partial_x \qty(\jmath'(q_0) \delta q + \tfrac{1}{2}\jmath''(q_0) \qty(\delta q)^2 + \cdots) = 0. \label{eq:expanded}
\end{equation}
To study correlation functions, we must also consider fluctuations away from the local equilibrium values. On physical grounds, this is well-described by an addition of hydrodynamical noise and phenomenological diffusion terms (End Matter). However, neither of the terms enter the asymptotic predictions below. 

To first order, Eq.~\eqref{eq:expanded} simply describes ballistic transport 
$\expval{\delta q(x, t)\delta
q(0,0)}\upd{c}_\mu \propto \delta(x - vt)$ with velocity $v = \jmath'(q_0)$. 
The spreading of this peak is quantitatively determined by the second-order term in \eqref{eq:expanded}, resulting in
\begin{align} \label{eq:kpz-correlation-asymptotic}
    \expval{\delta q(x, t)\delta q(0,0)}_\mu\upd{c} &= \frac{c_0}{(\lambda\ind{B}t)^{2/3}} f\ind{KPZ}\qty(\frac{x-vt}{(\lambda\ind{B}t)^{2/3}}),
\end{align}
with the superdiffusion constant $\lambda\ind{B} =
\sqrt{2c_0}\abs{\jmath''(q_0)}$ and normalization $c_0=
\expval{\hat{q}^2}_{\mu}-\expval{\hat{q}}_{\mu}^2$. The shape $f\ind{KPZ}$ is
exactly known and tabulated in Ref.~\onlinecite{Prhofer2004Exact}. As we show in the End Matter, for our model \eqref{eq:block_gate}, we
explicitly compute $\jmath(q) = \frac{1}{3}(\sqrt{16+9q^2}-2)$. At $\mu = 0$, we get
$c_0 = 4/9$ and $\lambda_B = 1/\sqrt{2}$. 

NLFHD predicts Eq.~\eqref{eq:kpz-correlation-asymptotic} as long as $\lambda\ind{B} \neq 0$. If $\lambda\ind{B} = 0$, NLFHD is dominated by the diffusive term and thus predicts the diffusive
scaling of correlations. This occurs, for instance, in the case of a gate that
conserves the total spin, or more generally when
$\hat{q}\upd{e}=\hat{q}\upd{o}$. In this case, it can be shown that
$\expval*{\hat{\jmath}}_\mu = 0$ and therefore $\jmath''(q) = 0$, implying
$\lambda\ind{B} = 0$. The superdiffusion constant can vanish also only at particular
$\mu$, such as in the example we present in the following section.

\begin{figure*}[ht]
    \centering
    \includegraphics[width=\textwidth]{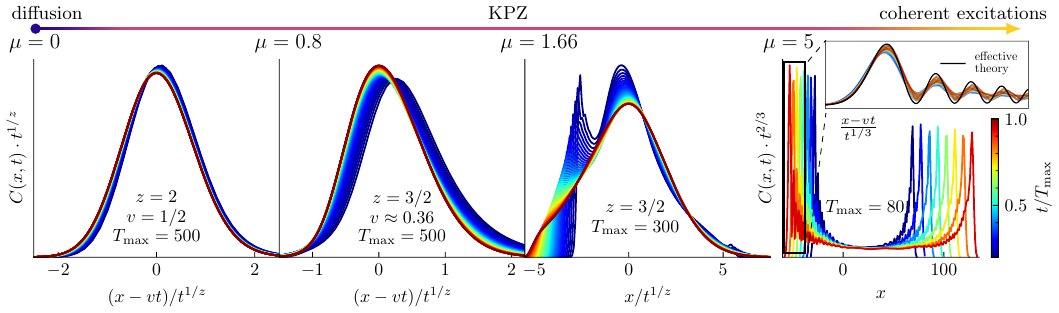}
    \caption{\textbf{Scaling of the charge-correlation profile in the three-site spin-$\tfrac{\boldsymbol 1}{\boldsymbol 2}$ circuit.} From left to right, the profiles are shown for $\mu=0,0.8,1.66$ and $5.0$, with the color indicating the evolution time. The data are shifted to the comoving coordinate $x-vt$ (at $\mu = 0,0.8)$ and computed with bond dimension $\chi = 300$. At $\mu=0$, rescaling the position and amplitude by $t^{1/2}$ produces a diffusive collapse over the accessible time window, whereas for intermediate $\mu = 0.8$ we see superdiffusive collapse when rescaling by $t^{2/3}$. For larger $\mu$, secondary peaks and oscillatory structures produce substantial finite-time deviations from the KPZ scaling. These features originate from dilute, long-lived ballistic quasiparticles and lead to an increasingly slow crossover toward the asymptotic hydrodynamic regime. The inset at $\mu = 5$ shows the predictions of the effective theory for the scaling [End Matter] and shape [SM~\onlinecite{SuppMat}: Eq.~\eqref{eq:effective_theory}] of the ballistic peak agreeing well with the numerics in the vicinity of the peak.} \label{fig:qubit_profiles}
    \vspace{-10pt}
\end{figure*}

 \begin{figure}[t]
    \centering
\includegraphics[width=0.49\textwidth]{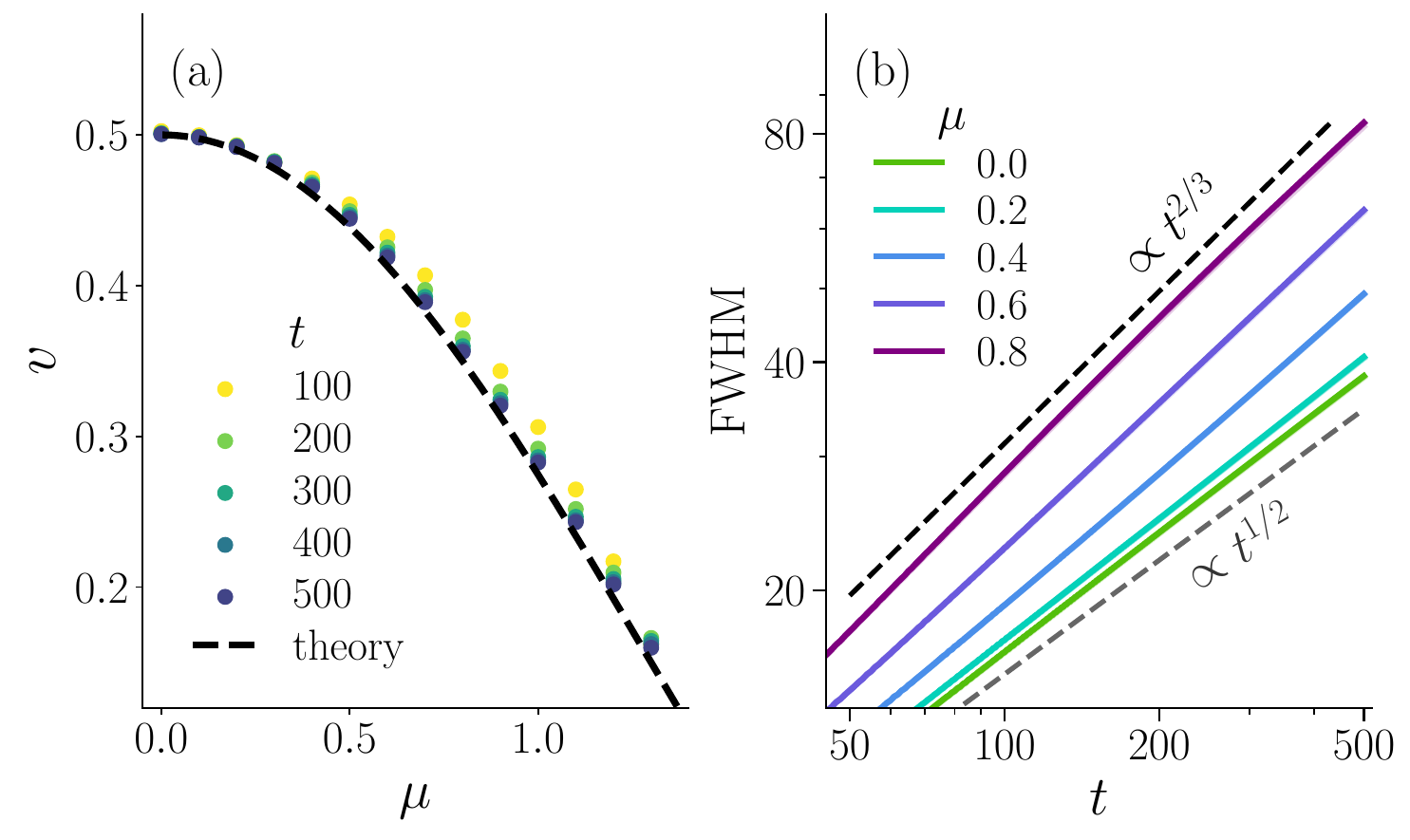}
    \caption{\textbf{Propagation and broadening of the charge-correlation in the three-site qubit circuit.} (a) Propagation velocity $v$ as a function of the chemical potential $\mu$, obtained from TEBD simulations with bond dimension $\chi=300$. Velocities are extracted from linear fits to the peak position over time windows $[t-10,t+10]$, with color indicating the central time $t$. The black dashed curve shows the hydrodynamic prediction $v(\mu)=\jmath'(q(\mu))$. (b) Log-log scaled plot of broadening of the propagating peak as a function of time for different values of $\mu$ estimated with the full width at half maximum (FWHM). Lighter-to-darker curves correspond to increasing bond dimension $\chi=100,200,$ and $300$. The dashed lines indicate diffusive broadening, $\mathrm{FWHM}\propto t^{1/2}$, and KPZ superdiffusive broadening $\mathrm{FWHM}\propto t^{2/3}$.}
    \vspace{-10pt}
    \label{fig:FWHM}
\end{figure}

\textit{\textbf{Spin-$\frac12$ chain with three-site gates}---}
Our construction can also be applied to other circuit architectures. To exemplify this, we present another model with superdiffusive transport, this time defined on a chain of spin-$\tfrac{1}{2}$ particles with gates that act on $r=3$ sites. A time step of unitary evolution is given by three layers of gates
\begin{equation}
    \hat{\mathbb{U}}:=\hat{\mathbb{U}}_2 \hat{\mathbb{U}}_1 \hat{\mathbb{U}}_0,~~~~\hat{\mathbb{U}}_i:=\prod_n \hat{U}_{i + 3n,i+ 3n+1,i + 3n+2}. \label{eq:3site_geom}
\end{equation}
Again, we restrict ourselves to charges that are sums of single site terms. In this setup, up to symmetries, we can classify all possible such charges and their corresponding models [SM~\onlinecite{SuppMat}]. Here, we consider the following conserved quantity 
\begin{align}
    \hat{Q} & = \sum_n \qty(\hat{S}^z_{3n-1} + 2\hat{S}^z_{3n} + \hat{S}^z_{3n+1}), \label{eq:charge_3site}
\end{align}
with charge density  $\hat{q}(x)=\hat{S}^z_{3x-1}+2\hat{S}^z_{3x} +\hat{S}^z_{3x+1}$ and $\hat{S}^z = \ketbra{0} - \ketbra{1}$. The generic model which conserves this charge is given by the three-site gate $\hat{U} = \hat{V}\hat{S}_{231\leftarrow 123}$, where
\begin{equation}
\hat V =
\resizebox{0.65\columnwidth}{!}{%
\begin{tikzpicture}[
    baseline=(M.center),
    blockline/.style={
        draw=black!35,
        dashed,
        line width=0.35pt,
        dash pattern=on 2pt off 1.3pt
    }
]
\matrix (M) [
    matrix of math nodes,
    ampersand replacement=\&,
    left delimiter={(},
    right delimiter={)},
    nodes={
        anchor=center,
        inner xsep=1.7pt,
        inner ysep=1.7pt
    },
    column sep=0pt,
    row sep=0pt
] {
    {}
    \& \bra{001}
    \& \bra{100}
    \& \bra{010}
    \& \bra{101}
    \& \bra{011}
    \& \bra{110}
    \& \bra{000}
    \& \bra{111} \\
    \ket{001}
    \& \ast \& \ast \& {} \& {} \& {} \& {} \& {} \& {} \\
    \ket{100}
    \& \ast \& \ast \& {} \& {} \& {} \& {} \& {} \& {} \\
    \ket{010}
    \& {} \& {} \& \ast \& \ast \& {} \& {} \& {} \& {} \\
    \ket{101}
    \& {} \& {} \& \ast \& \ast \& {} \& {} \& {} \& {} \\
    \ket{011}
    \& {} \& {} \& {} \& {} \& \ast \& \ast \& {} \& {} \\
    \ket{110}
    \& {} \& {} \& {} \& {} \& \ast \& \ast \& {} \& {} \\
    \ket{000}
    \& {} \& {} \& {} \& {} \& {} \& {} \& \ast \& {} \\
    \ket{111}
    \& {} \& {} \& {} \& {} \& {} \& {} \& {} \& \ast \\
};
\coordinate (xleft) at (M-1-2.west);
\coordinate (xmiddleone) at
    ($(M-1-3.east)!0.5!(M-1-4.west)$);
\coordinate (xmiddletwo) at
    ($(M-1-5.east)!0.5!(M-1-6.west)$);
\coordinate (xmiddlethree) at
    ($(M-1-7.east)!0.5!(M-1-8.west)$);
\coordinate (xmiddlefour) at
    ($(M-1-8.east)!0.5!(M-1-9.west)$);
\coordinate (xright) at (M-1-9.east);
\coordinate (ytop) at
    ($(M-1-2.south)!0.5!(M-2-2.north)$);
\coordinate (ymiddleone) at
    ($(M-3-2.south)!0.5!(M-4-2.north)$);
\coordinate (ymiddletwo) at
    ($(M-5-2.south)!0.5!(M-6-2.north)$);
\coordinate (ymiddlethree) at
    ($(M-7-2.south)!0.5!(M-8-2.north)$);
\coordinate (ymiddlefour) at
    ($(M-8-2.south)!0.5!(M-9-2.north)$);
\coordinate (ybottom) at (M-9-1.south);
\draw[blockline]
    (xleft |- ytop)
    --
    (xleft |- ybottom);
\draw[blockline]
    (xmiddleone |- ytop)
    --
    (xmiddleone |- ybottom);
\draw[blockline]
    (xmiddletwo |- ytop)
    --
    (xmiddletwo |- ybottom);
\draw[blockline]
    (xmiddlethree |- ytop)
    --
    (xmiddlethree |- ybottom);
\draw[blockline]
    (xmiddlefour |- ytop)
    --
    (xmiddlefour |- ybottom);
\draw[blockline]
    (xright |- ytop)
    --
    (xright |- ybottom);
\draw[blockline]
    (xleft |- ytop)
    --
    (xright |- ytop);
\draw[blockline]
    (xleft |- ymiddleone)
    --
    (xright |- ymiddleone);
\draw[blockline]
    (xleft |- ymiddletwo)
    --
    (xright |- ymiddletwo);
\draw[blockline]
    (xleft |- ymiddlethree)
    --
    (xright |- ymiddlethree);
\draw[blockline]
    (xleft |- ymiddlefour)
    --
    (xright |- ymiddlefour);
\draw[blockline]
    (xleft |- ybottom)
    --
    (xright |- ybottom);
\end{tikzpicture}%
}
\label{eq:block_gate_3site}
\end{equation}
and $\hat{S}_{231 \leftarrow 123} = \sum_{a,b,c} \ketbra{cab}{abc}$. Again, each of the $*$ blocks can be a generic unitary matrix. For the charge and current in equilibrium, we obtain [SM~\onlinecite{SuppMat}]
\begin{equation}
    q(\mu)=\tanh \mu+\tanh \frac{\mu}{2},~~~
    \jmath(\mu)=\tanh \mu-\frac{1}{2}\tanh \frac{\mu}{2}.
\end{equation}
Interestingly, at $\mu=0$, the nonlinear term vanishes $\jmath''(q(0))=0$, and we hence find diffusive spreading with velocity $v(0)=\jmath'(q(0)) =1/2$. As shown in Fig.~\ref{fig:FWHM}(a), the hydrodynamic velocity prediction agrees well with the velocity extracted from the TEBD simulations, also over the entire range of chemical potentials considered. The diffusive broadening of the propagating peak is likewise consistent with NLFHD. For generic  $\mu > 0$, $\jmath''(q(\mu))\neq0$, and NLFHD instead predicts KPZ broadening $z = 3/2$. The numerical results in Fig.~\ref{fig:FWHM}(b) agree with a crossover between these two regimes at finite times.

As shown in Fig.~\ref{fig:qubit_profiles}, we observe that the convergence to the exact KPZ profile is much slower for the spin-$\frac{1}{2}$ model compared to the spin-$1$ case \eqref{eq:block_gate}. This is due to slowly decaying ballistic peaks of `gliders', or solitons, described by an effective low-charge quantum integrable theory which we can obtain for $\mu \to \infty$. At large $\mu$, all states of the ensemble can be interpreted as few-charge excitations over the unique ground state $\ket{\cdots 00000 \cdots}$. Similar to the low-energy excitations of Hamiltonian models, single excitations above this ground state are ballistic quasiparticles whose quasimomentum is conserved [End Matter, SM~\onlinecite{SuppMat}]. In two quasiparticle scatterings, both quasimomentum and quasienergy are conserved, and hence the system is approximately integrable; integrability is only broken upon three or more particle scattering. Therefore, for large $\mu$, where the density of excitations is low, integrability and its associated ballistic transport survive for long times.

Thus, the approach to the KPZ profile is slower for larger $\mu$. We plot our prediction from the above effective theory  for the shape of the solitonic peak [End Matter: Eq.~\eqref{eq:airy}] in the inset of Fig.~\ref{fig:qubit_profiles}, demonstrating good agreement with numerics for large $\mu$ in the vicinity of the peak. A detailed study of this slow integrability breaking is beyond the scope of this letter, but is an interesting problem for future works, especially because the crossover from ballistic to diffusive or superdiffusive transport due to weak integrability breaking \cite{Weak_Surace,Weak_Us,Weak_me,vanovac2026weakintegrabilitybreakingperturbations} is an area of active study~\cite{8232-shpj,PhysRevB.101.180302,SciPostPhys.18.6.186,Bastianello_2021,PhysRevLett.127.130601}.

Let us remark that an analogous integrable model at low charges does not exist in the spin-$1$ model \eqref{eq:block_gate}. This is simply because for $\mu \to \infty$, $e^{\mu\hat{Q}}$ is a mixed state instead of a pure state. Hence, the ``low-charge'' excitations are not quantum coherent quasiparticles but instead should be viewed as incoherent thermal excitations. Therefore, the approach to the asymptotic KPZ profile is significantly faster.

\textit{\textbf{Discussion}---} We have shown that, unlike chaotic
locally-interacting Hamiltonian spin chains, chaotic quantum circuits can
transport charge superdiffusively. Our results are backed by an effective
theory based on nonlinear fluctuating hydrodynamics (NLFHD) and extensive
tensor-network simulations. While for the two-site spin-$1$ circuit, our numerical
results agree remarkably well with the parameter-free NLFHD prediction, we
observe a long crossover in the three-site spin-$\frac12$ case. We attribute this long
crossover to dilute, long-lived quantum quasiparticles, which we explicitly
identify in the large-$\mu$ limit.

Although classical circuits may also support gliders or solitons, the coherent
quasiparticle character of the found excitations is intrinsically quantum. They
are as such expected to be particularly relevant for the description of purely quantum
quantities, for instance out-of-time order correlators (OTOCs). It would be interesting
to develop a theory for these long-lived particles and to see whether eventually
the description becomes purely classical or whether the quantum effects remain
to be relevant.

More broadly, our results show that KPZ-like superdiffusion in quantum systems
is not confined to fine-tuned integrable models, but can also emerge from fully
chaotic dynamics. In contrast to integrable models, NLFHD predicts the transport
to genuinely belong to the KPZ universality class. Probing this in more detail (for instance, the fluctuations of the currents)
is a compelling avenue to explore that would require and improvement in either
numerical methods or a more analytically tractable model.

Due to the recent advances in quantum computing, both the long-lived quantum
quasiparticles and the emergence of KPZ universality in the proposed models can
also be studied experimentally. They provide an especially interesting target for
quantum simulations in the large $\mu$ regime where the quasiparticles dominate
and the tensor-network methods become less feasible. A generalization of our
construction with multiple charges, including quasilocal ones \cite{Quasicharges}, which would give us access to other
universality classes \cite{Popkov_Fibonacci}, is also an interesting open question that would further
cement quantum circuits as a flexible platform for realizing NLFHD in genuinely
quantum many-body systems.

\section{Acknowledgements} 
We would like to thank Marko Ljubotina for suggesting an important improvement
to our numerical methods. We also thank Pavel Kos, Frank Pollmann and
Mari Carmen Bañuls for useful discussions. We thank Luka Paljk for helpful comments on the manuscript.  RS acknowledges support from ERC Advanced
Grant No.~101096208—QUEST, and Research Programme P1-0402 of Slovenian Research
and Innovation Agency (ARIS). UD acknowledges Grants No.~N1-0504,
No.~J1-70049, and No.~P1-0402 from Slovenian Research Agency (ARIS). SWPK is
supported by Engineering and Physical Sciences Research Council (EPSRC) DTP
International Studentship Grant Ref.~No.~EP/W524475/1. FH has received support
under the Major Research Program of PSL Research University "Statistical Physics
and Mathematics" launched by PSL Research University and implemented by ANR with
the references ANR-10-IDEX-0001. 

\newpage
\section{End matter}

\paragraph{\textit{\textbf{Current in the spin-$\boldsymbol 1$ model.}}---}
The microscopic continuity equation is given by
\begin{align}
\hat{\mathbb{U}}^{\dagger}\hat{q}(x)~\hat{\mathbb{U}} - \hat{q}(x) &= \hat{\jmath}(x-\tfrac12) - \hat{\jmath}(x+\tfrac12). 
\label{eq:contin}
\end{align}
Let us simplify the LHS using  $\hat{U}^{\dagger} \hat{q} ~\hat{U}=\hat{S} ~\hat{q} ~\hat{S}$ from the main text~\eqref{VqVq}
\begin{align}
\begin{aligned}
&~~~\picR[scale=0.025][-0.25em]{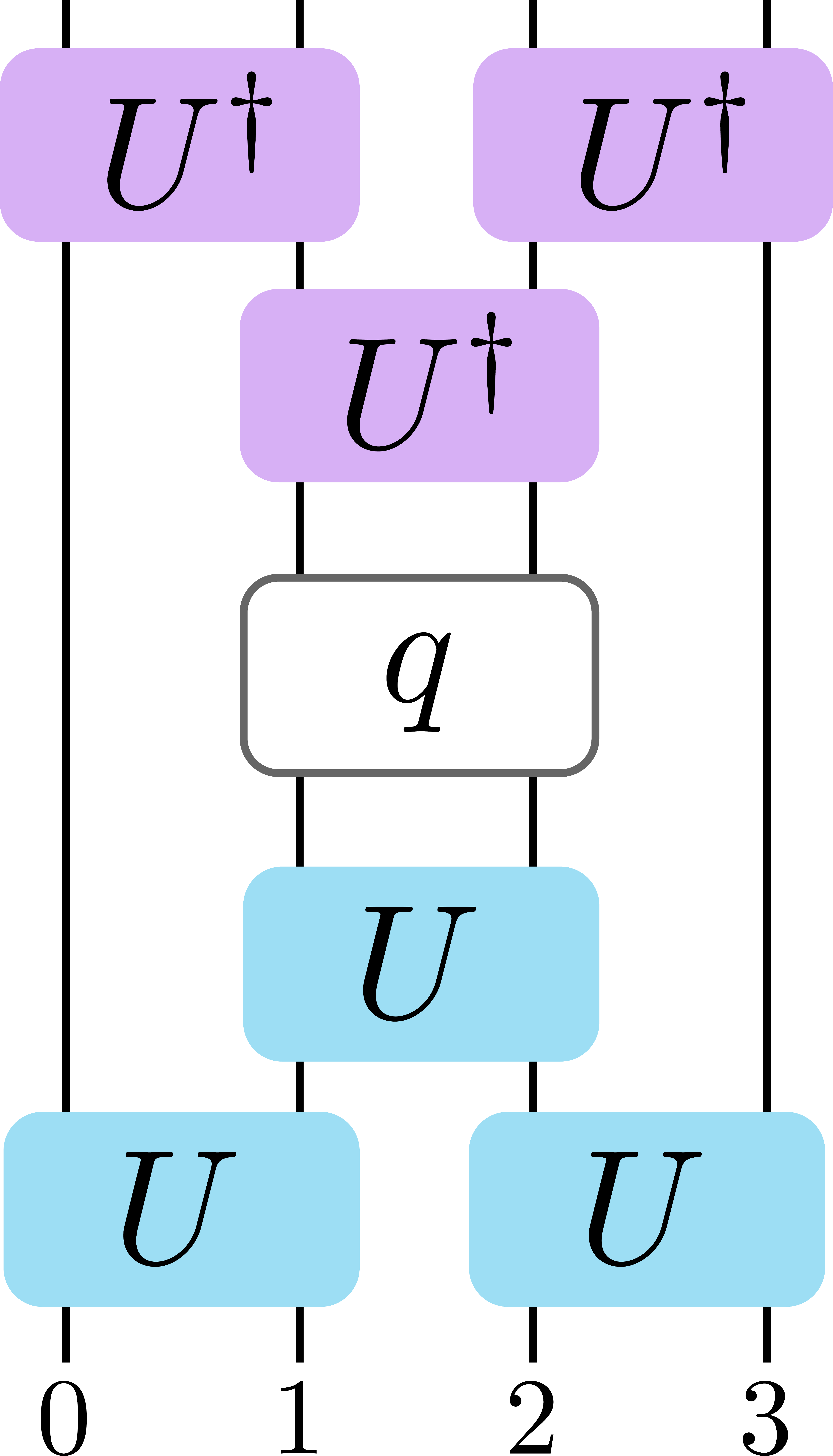}~~=~~\picR[scale=0.025][-0.25em]{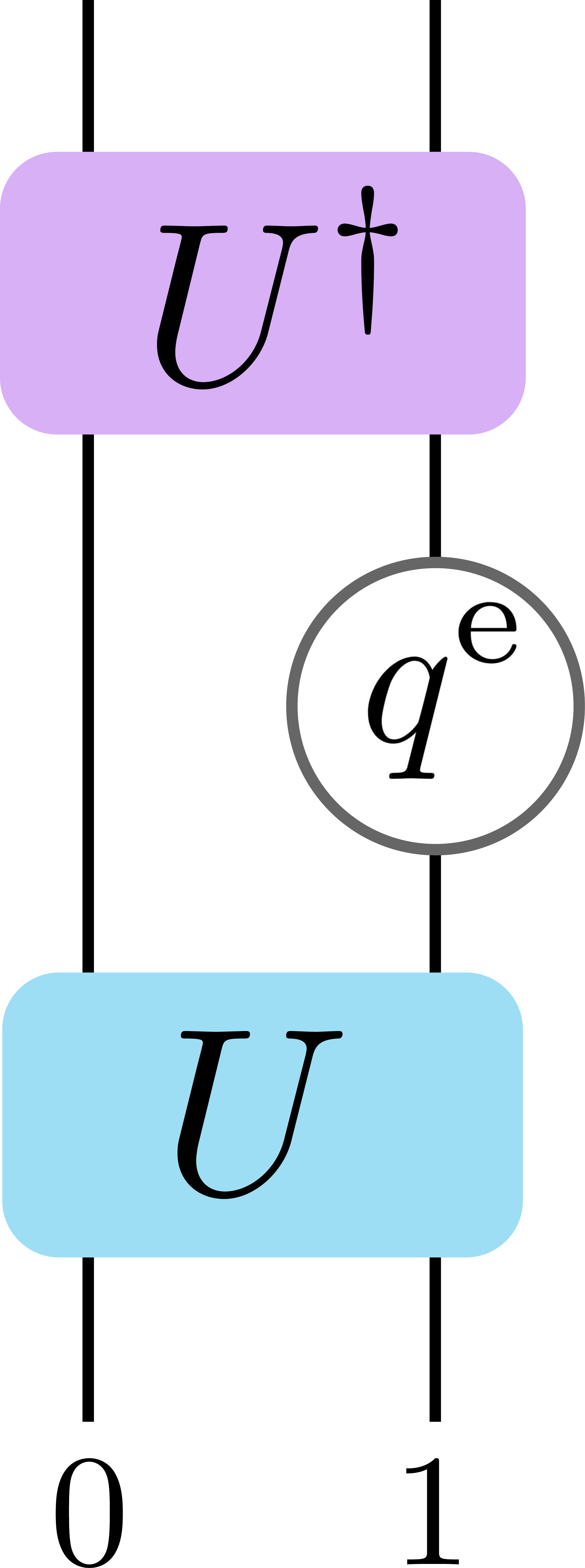}~+~\picR[scale=0.025][-0.25em]{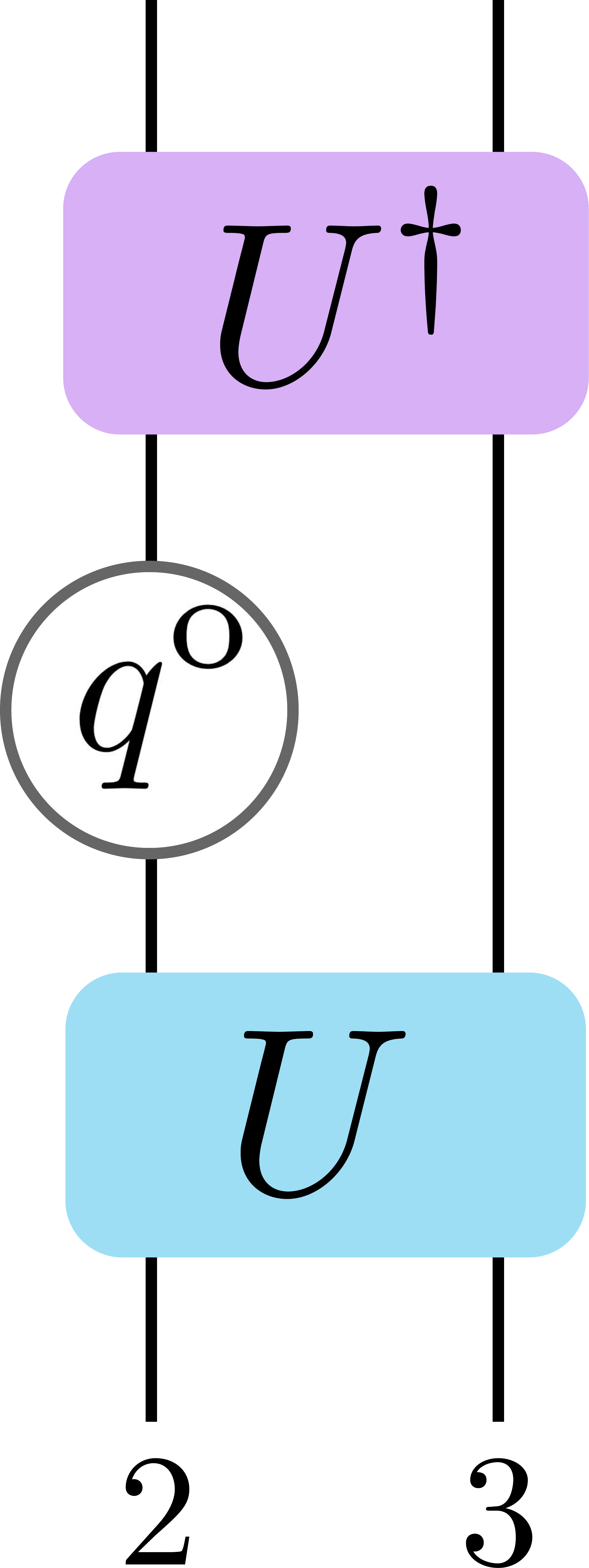}~.
\end{aligned}
\end{align}
The RHS can be simplified by using the identity  $\hat{q}^\mathrm{o} \otimes \hat I= \hat{q}- \hat I\otimes \hat{q}^\mathrm{e}$,
\begin{align}
\begin{aligned}
\picR[scale=0.025][-0.3em]{UqeU}-\picR[scale=0.025][-0.3em]{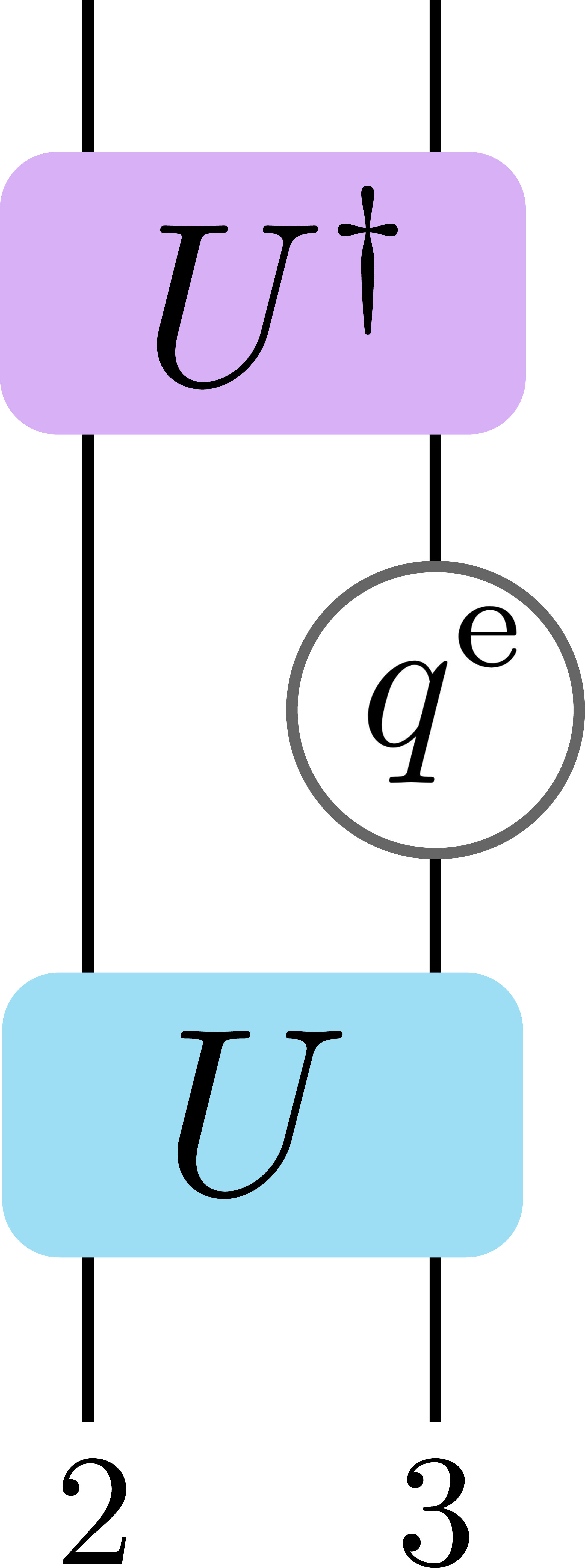}+\picR[scale=0.025][-0.3em]{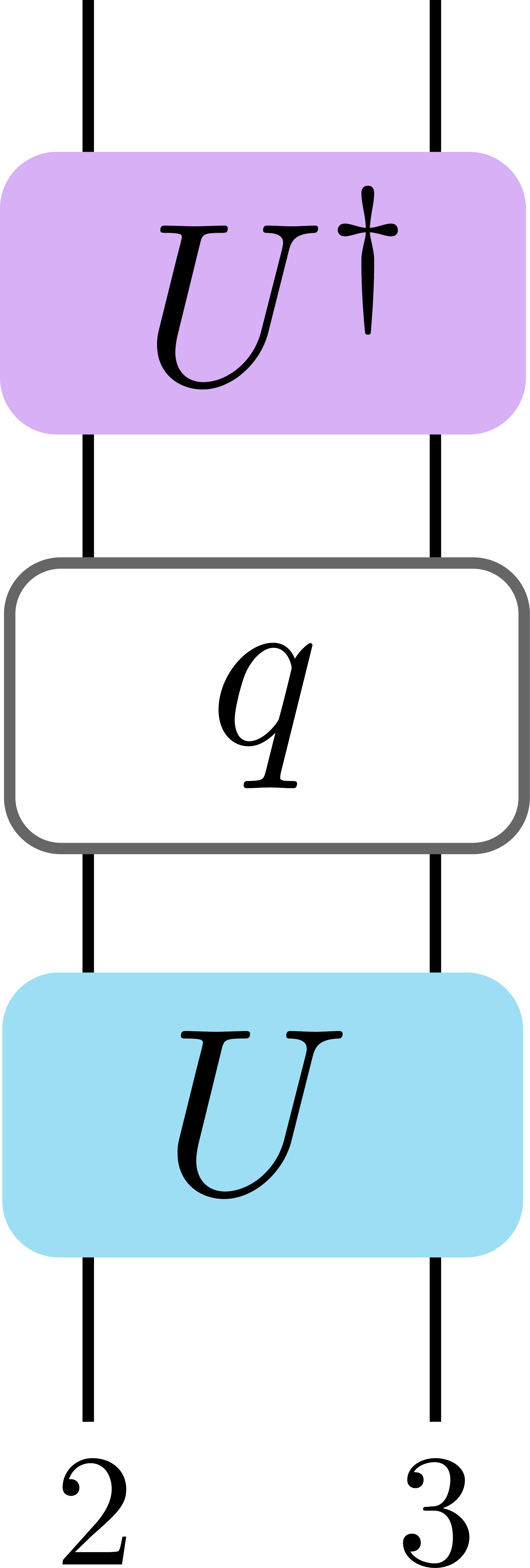}~=~\picR[scale=0.025][-0.3em]{UqeU}-\picR[scale=0.025][-0.3em]{Uqe23U}+\picR[scale=0.025][-0.3em]{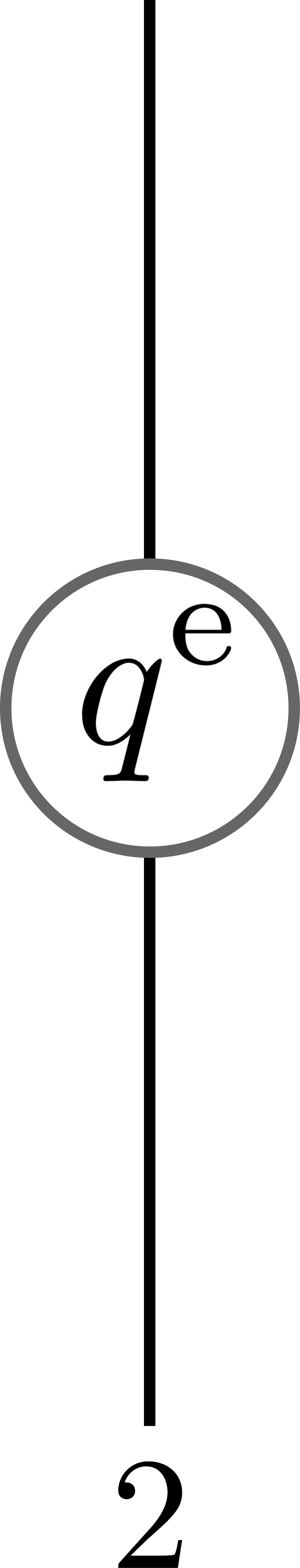}+\picR[scale=0.025][-0.3em]{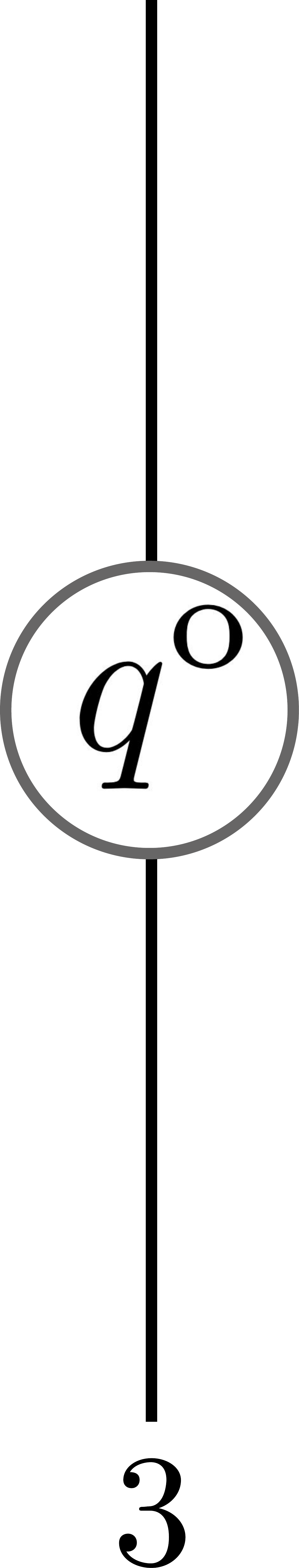}~,\nonumber
\end{aligned}
\end{align}
where we used again that $\hat{U}^{\dagger} \hat{q}~ \hat{U}=\hat{S}~\hat{q} ~\hat{S}$. Finally, we obtain Eq. \eqref{eq:contin} written in the diagrammatic form 
\begin{align}
\begin{aligned}
~\picR[scale=0.025][-0.3em]{UUqUU}~-~~\picR[scale=0.025][-0.3em]{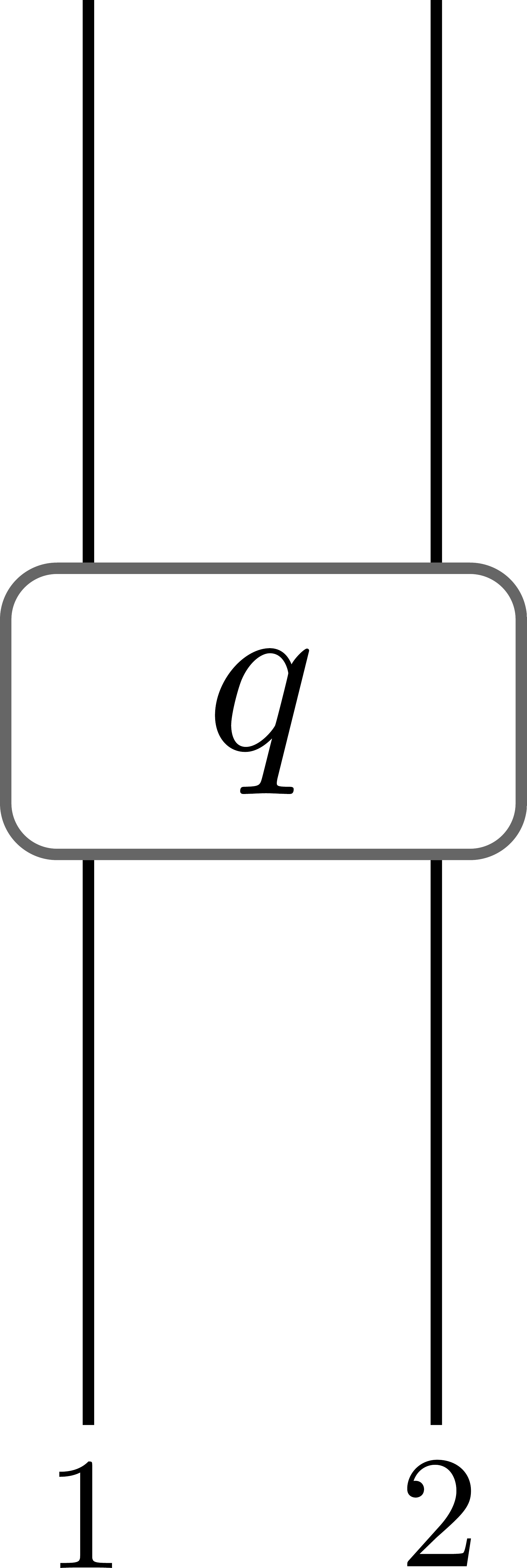}~=~\picR[scale=0.025][-0.3em]{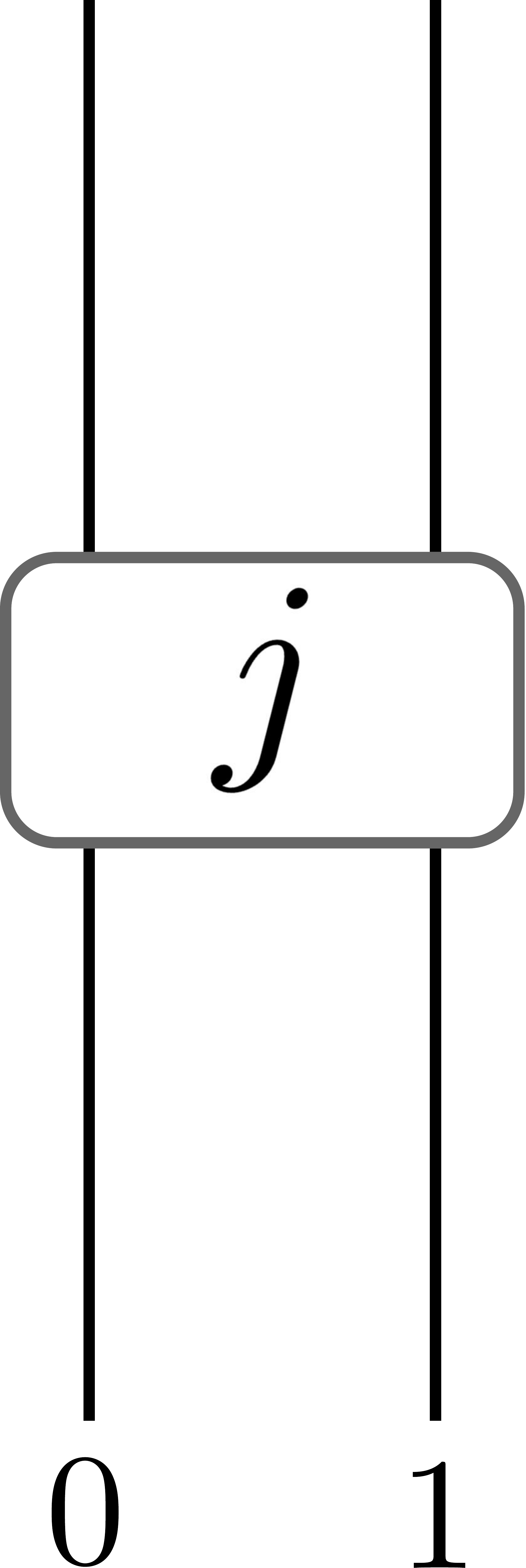}~-~\picR[scale=0.025][-0.3em]{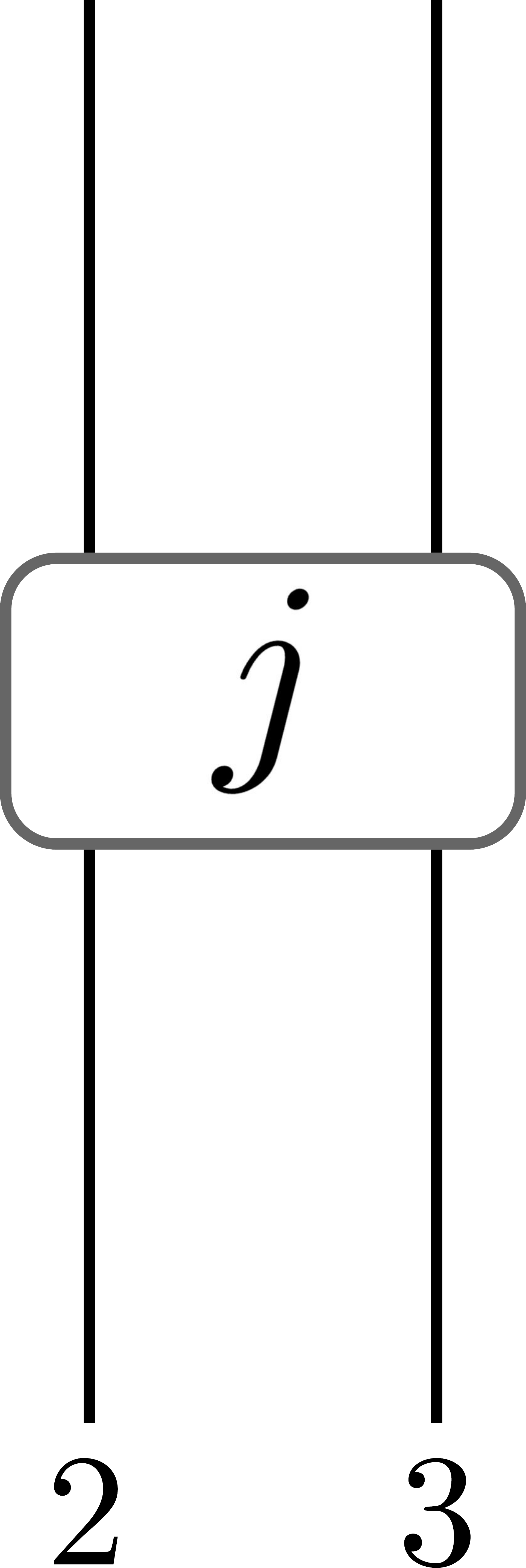}~,
\end{aligned}
\end{align}
where     
\begin{align}
\begin{aligned}
&~~~ \picR[scale=0.026][-0.0em]{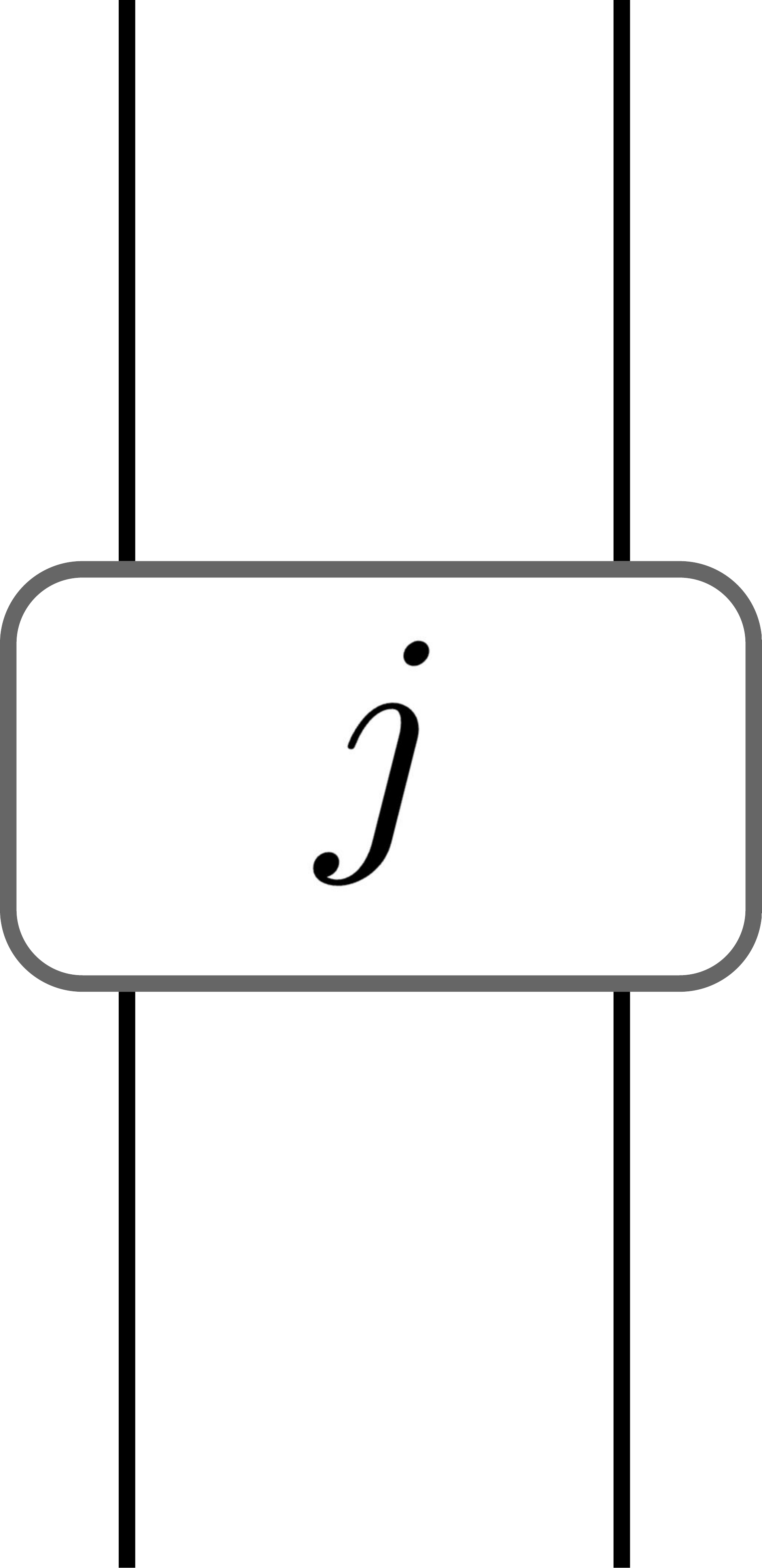}~:=~ \, \picR[scale=0.027][-0.0em]{UqeUn}~-~~\picR[scale=0.026][-0.0em]{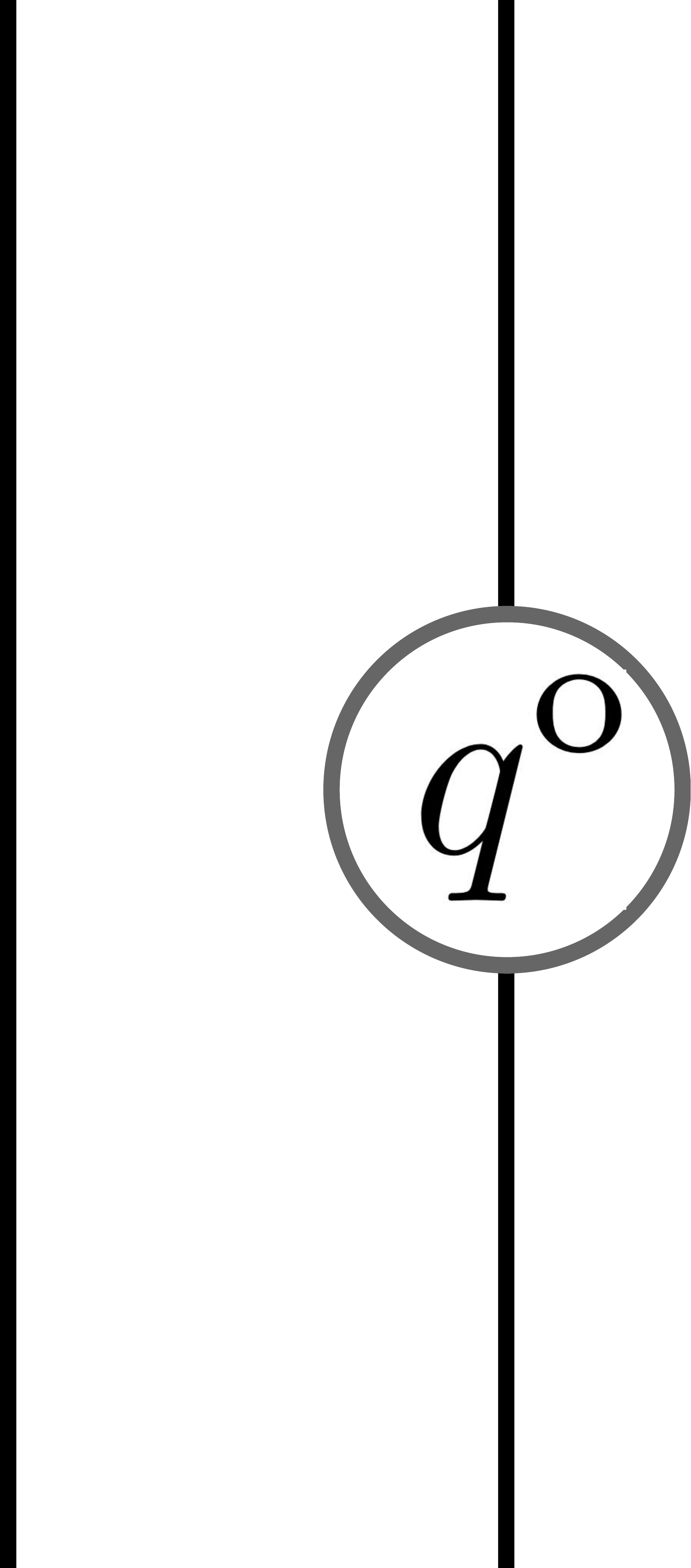}~.
\end{aligned}
\label{eq:jdef}
\end{align}

{\textit{\textbf{Equilibrium expectation values in the spin-$\boldsymbol 1$ model.}}--- We now evaluate the expectation value of the current and the charge in the
equilibrium state. Let us first define the one-site density matrices
\begin{equation}
  \hat\rho_\mu^a
  := \frac{e^{\mu\hat q^a}}{\operatorname{tr}\left[e^{\mu\hat q^a}\right]},
  \qquad a\in\{\mathrm e,\mathrm o\}
  \label{eq:local-gibbs-states}
\end{equation}
and the equilibrium states on the even-odd pair and the odd-even pair (respectively) as
\begin{equation}
  \hat\rho_\mu^{\mathrm{eo}}:=\hat\rho_\mu^{\mathrm e}\otimes\hat\rho_\mu^{\mathrm o}, \quad \hat\rho_\mu^{\mathrm{oe}}:=\hat\rho_\mu^\mathrm{o}\otimes\hat\rho_\mu^\mathrm{e}.
\end{equation}
Upon exchanging the two sites, we have
\begin{equation}
  \hat\rho_\mu^{\mathrm{oe}}
  = \hat S\hat\rho_\mu^{\mathrm{eo}}\hat S
  = \frac{e^{\mu\hat q}}{\operatorname{tr}\left[e^{\mu \hat{q}}\right]}.
  \label{eq:swapped-gibbs-state}
\end{equation}
From  $\hat{U}^{\dagger} \hat{q} \hat{U}=\hat{S} \hat{q} \hat{S}$ it follows that $\hat{U}^\dagger \hat{\rho}^\mathrm{oe} \hat{U}=\hat{S}
\hat{\rho}^\mathrm{oe}\hat{S}=\hat{\rho}^\mathrm{eo}$, which is equivalent to  $\hat{U} \hat{\rho}^\mathrm{eo}_\mu\hat{U}^\dagger=\hat{\rho}_\mu^\mathrm{oe}$. Consequently,
\begin{align}
 &\left\langle
   \hat U^\dagger
   \bigl(\hat{I}\otimes\hat q^{\mathrm e}\bigr)
   \hat U
  \right\rangle_\mu
  =
  \operatorname{tr}\!\left[\hat{U}
    \hat\rho_\mu^{\mathrm{eo}}
    \hat{U}^{\dagger}
    \bigl(\hat{I}\otimes\hat q^{\mathrm e}\bigr)
  \right]
  \nonumber\\
 &=
  \operatorname{tr}\!\left[
    \hat\rho_\mu^{\mathrm{oe}}
    \bigl(\hat{I}\otimes\hat q^{\mathrm e}\bigr)
  \right]
  = \left\langle\hat q^{\mathrm e}\right\rangle_\mu .
  \label{eq:equilibrium-current-trace}
\end{align}
Now, using Eq.~\eqref{eq:jdef}, we are able to find the equilibrium value of the current 
\begin{equation}
  \jmath(\mu)
  := \left\langle\hat \jmath \right\rangle_\mu
  =  \left\langle
   \hat U^\dagger
   \bigl(\hat{I}\otimes\hat q^{\mathrm e}\bigr)
   \hat U
  \right\rangle_\mu - \left\langle\hat q^{\mathrm o}\right\rangle_\mu
  = \left\langle\hat q^{\mathrm e}\right\rangle_\mu
    -\left\langle\hat q^{\mathrm o}\right\rangle_\mu .
  \label{eq:equilibrium-current}
\end{equation}
The equilibrium charge density is, similarly,
\begin{equation}
  q(\mu)
  = \left\langle\hat q^{\mathrm e}\right\rangle_\mu
    +\left\langle\hat q^{\mathrm o}\right\rangle_\mu .
  \label{eq:equilibrium-charge}
\end{equation}

For the spin-$1$ model \eqref{eq:block_gate},
$\hat q^{\mathrm e}=(\hat{S}^z_n)^2 $ and
$\hat q^{\mathrm o}=-(\hat{S}^z_n)^2$. So the equilibrium expectations are
\begin{equation}
  \left\langle\hat q^{\mathrm e}\right\rangle_\mu
  = \frac{e^\mu}{e^\mu+2},
  \qquad
  \left\langle\hat q^{\mathrm o}\right\rangle_\mu
  = -\frac{1}{1+2e^\mu}.
\end{equation}
It follows that
\begin{align}
  q(\mu)
  &= \frac{e^\mu}{e^{\mu}+2}-\frac{1}{1+2e^\mu},
  \label{eq:qutrit-charge-mu}\\
  \jmath(\mu)
  &= \frac{e^\mu}{e^\mu+2}+\frac{1}{1+2e^\mu}.
  \label{eq:qutrit-current-mu}
\end{align}

\textit{\textbf{Details of the coarse-grained description}---}
The idea behind a hydrodynamic coarse-grained description is as follows: imagine running a large circuit of size $L\gg 1$ for a long time $1 \ll T \ll L$. Then, because the evolution of the circuit is chaotic, we expect that it will locally equilibrate to a Gibbs state (i.e., a maximum entropy state) of the form $\frac{1}{Z(\mu)} e^{\mu \hat{Q}}$. However, since the system is much larger than the light-cone, equilibration cannot yet occur globally. Instead, far away local regions equilibrate to different Gibbs states. Therefore, a natural assumption for the state of the system is a so-called local Gibbs state
\begin{align}
    \hat{\rho} &= \frac{1}{Z} \exp\qty(\sum_{x} \mu \left(\tfrac{x}{L}, \tfrac{t}{L}\right) \hat{q}(x)).
\end{align}
Here $\mu(x,t)$ is a slowly varying ``chemical potential'' characterizing the local Gibbs state evolving in time.

Next, we need to understand the evolution of such local Gibbs states. To this end, consider the continuity equation 
\begin{align}
    \hat{q}(x,t+1) - \hat{q}(x,t) &= \hat{\jmath} \left(x-\tfrac12,t\right) - \hat{\jmath}\left(x+\tfrac12,t\right).  
\end{align}
The average charge and current in the local Gibbs state are given by \eqref{eq:qutrit-charge-mu} and \eqref{eq:qutrit-current-mu}
\begin{align}
    q(x,t) &= q(\mu(x,t)) & \jmath(x,t) &= \jmath(\mu(x,t)).
\end{align}
Instead of thinking of the current $\jmath$ as a function $\jmath(\mu)$, we can also think of it as a function of $q$ as $\jmath(q) = \jmath(\mu(q))$, where $\mu(q)$ is the ``chemical potential'' for which $\expval{\hat{q}}_{\mu(q)} = q$. This way, we obtain
\begin{align}
    \jmath(q) &=\frac{\sqrt{16+9q^2}-2}{3}.
\end{align}
With that, the continuity equation becomes the so-called Euler hydrodynamic equation
\begin{align}
    \partial_t q(x,t) + \partial_x \jmath(q(x,t)) &= 0.\label{eq:euler_app}
\end{align}

This equation is the backbone of hydrodynamics and describes ballistic transport on very large scales. According to the theory of nonlinear fluctuating hydrodynamics~\cite{spohn2014}, the next-order effects are diffusion and noise, linked together by the fluctuation- dissipation theorem~\cite{spohn2014}, giving an equation of the form
\begin{align}
    \partial_t q(x, t) + \partial_x \jmath(q) &= \partial_x(D(q)\partial_x q + B(q)\xi(x, t)), 
\end{align}
where $D$ is the diffusion constant, $\xi(x, t)$ is the white noise $\expval{\xi(x, t)} = 0$, $\expval{\xi(x, t)\xi(y, s)} =\delta(x-y)\delta(t-s)$ and $B$ characterizes the noise strength. $D$ and $B$ are related via the fluctuation-dissipation relation $2DC=B^2$. It is an open problem to fix the values of $D$ and $B$ from first principles~\cite{spohn2014,PhysRevLett.127.010601,2z9s-5d7w,koterle2026effectivefieldtheoriesnonlinear}; however, they do not affect the asymptotic result for correlation functions.

To apply this theory to correlation functions, we expand $q(x, t) = q_0 + \delta q(x, t)$ around an equilibrium value
\begin{align}
    \partial_t \delta q + v(q_0) \partial_x\delta q+\tfrac{1} {2}\jmath''(q_0) \partial_x \delta q^2 - D\partial_x^2 \delta q + B\partial_x \xi &= 0.\label{equ:burgers}
\end{align}
Using the Galilei invariance, we can set $v(q_0)=0$. We have obtained the stochastic Burgers equation which can be mapped to the KPZ equation
\begin{align}
    \partial_t h + \tfrac{1}{2}\jmath''(q_0) \left(\partial_x h\right)^2 - D\partial_x^2 h + B\xi = 0
\end{align}
using $\delta q(x, t) = \partial_x h(x, t)$. We can thus obtain the asymptotics of \eqref{equ:burgers} via the known asymptotics of the KPZ equation~\cite{Prhofer2004Exact}, yielding \eqref{eq:kpz-correlation-asymptotic}.

\textit{\textbf{``Low charge'' excitations of three-site qubit gates}---} Here, we briefly explain how to obtain the theoretical prediction for large $\mu$ in Fig. \ref{fig:qubit_profiles}. More details can be found in the SM~\cite{SuppMat}.
We start with the observation that in model \eqref{eq:block_gate_3site}, for $\mu \to \infty$, the Gibbs state becomes a pure state $\hat{\rho}_\mu \to \ket{\Omega}\bra{\Omega}$, where $\ket{\Omega} = \ket{\cdots 000000\cdots}$. For $\mu$ large but finite, we thus expect that only states with a low charge are activated. This is completely analogous to a Hamiltonian system, where for large inverse temperature $\beta$ (i.e., low temperature) only low energy excitations are present. In model \eqref{eq:block_gate_3site}, the sector with the lowest charge excitation $1$ is spanned by the states $\ket{3j+1},\ket{3j+2}$, where $\ket{j}$ is the state with $1$ at site $j$ and $0$ elsewhere. Assuming that $\mu$ is so large that we can neglect all other sectors, we can exactly solve the evolution. In particular, we can find the eigenstates of $\hat{\mathbb{U}}$ in this subspace: since $\hat{\mathbb{U}}$ is translation invariant in momentum space, the problem reduces to diagonalizing a $2\times 2$ matrix~[SM~\onlinecite{SuppMat}] for each momentum $k$. In the diagonal basis, the problem is equivalent to two independent quantum particles (labeled by $\pm$) with dispersion relations $\varepsilon^\pm_k$ that can be extracted from the eigenvalues. The time-evolution of the wavefunction is then given by oscillatory integrals, which for large times can be approximated by the stationary phase approximation, yielding a decay $\sim t^{-1}$ of the correlation function. However, the dispersion relation is such that for certain $k=k^\pm$ we have $\partial_k^2 \varepsilon^\pm_k=0$. At those points the stationary phase approximation cannot be applied. Instead, the correlation function decays slower $\sim t^{-2/3}$. Asymptotically, it becomes an Airy shape~[SM~\onlinecite{SuppMat}]:
\begin{equation}
    \expval{\hat{q}(x,t)\hat{q}(0,0)}\upd{c} 
    \approx e^{-\mu}\sum_{\sigma=\pm}\qty(\frac{\alpha_\sigma}{t})^{\tfrac{2}{3}}
    \mathrm{Ai}^2\left(-\sqrt[3]{\frac{\alpha_\sigma}{t}}(x-v^\sigma t)\right), \label{eq:airy}
\end{equation}
where $\alpha_\pm = 18/\partial_{k}^3\varepsilon^\pm_{k^\pm}$. The scaling predicted by this effective theory is in good agreement with that of numerical simulations for $\mu=5$, as shown in Fig. \ref{fig:qubit_profiles}.

\bibliography{bibliography}
\clearpage

\newpage
\onecolumngrid
\clearpage
\onecolumngrid
\appendix
\setcounter{secnumdepth}{3} 
\renewcommand{\thesection}{\Alph{section}}
\counterwithout{equation}{section}
\setcounter{equation}{0}
\renewcommand{\theequation}{S\arabic{equation}}
\renewcommand{\theHequation}{supplement.\arabic{equation}}

\makeatletter
\def\@sectioncntformat#1{\csname the#1\endcsname.}
\def\@hangfrom@section#1#2#3{#1#2\quad\MakeTextUppercase{#3}}
\makeatother
\begin{center}
	{\Large Supplemental Material for ``Kardar–Parisi–Zhang superdiffusion in chaotic quantum circuits''}
\end{center}

\SMcontents

\SMsection{Explicit models used for simulations}
Here we list the specific gates used for simulations in the main text. The spin-1 two-site gate is given by $\hat{U}=\hat{V}\hat{S}$, where

\begin{equation}
\hat V =
\resizebox{0.96\columnwidth}{!}{%
\begin{tikzpicture}[
    baseline=(M.center),
    blockline/.style={
        draw=black!35,
        dashed,
        line width=0.35pt,
        dash pattern=on 2pt off 1.3pt
    }
]
\matrix (M) [
    matrix of math nodes,
    ampersand replacement=\&,
    left delimiter={(},
    right delimiter={)},
    nodes={
        anchor=center,
        inner xsep=1.7pt,
        inner ysep=1.7pt
    },
    column sep=0pt,
    row sep=0pt
] {
    {}
    \& \bra{0\small{\uparrow}}
    \& \bra{0\small{\downarrow}}
    \& \bra{\small{\uparrow} 0}
    \& \bra{\small{\downarrow} 0}
    \& \bra{00}
    \& \bra{\downarrow\downarrow}
    \& \bra{\downarrow\uparrow}
    \& \bra{\uparrow\downarrow}
    \& \bra{\uparrow\uparrow} \\
    \ket{0\small{\uparrow}}
    \& 0.46403 -  0.45271  i \& -0.73729 + 0.19009 i \& {} \& {} \& {} \& {} \& {} \& {} \& {} \\
    \ket{0\small{\downarrow}}
    \& -0.5147 - 0.56108i \& -0.14248 - 0.63243 i  \& {} \& {} \& {} \& {} \& {} \& {} \& {} \\
    \ket{\small{\uparrow} 0}
    \& {} \& {} \& 0.18384 + 0.19805 i \& -0.91735 + 0.29233 i \& {} \& {} \& {} \& {} \& {} \\
    \ket{\small{\downarrow} 0}
    \& {} \& {} \& -0.95403 - 0.12967 i \& -0.14691 + 0.2268 i \& {} \& {} \& {} \& {} \& {} \\
    \ket{00}
    \& {} \& {} \& {} \& {} \& 0.30367 - 0.00927 i \& 0.0612 + 0.40908 i \& -0.0176 + 0.37929 i \& 0.16125 + 0.56413 i \& -0.04441 - 0.49621 i \\
    \ket{\downarrow\downarrow}
    \& {} \& {} \& {} \& {} \& -0.11164 + 0.12646 i \& 0.63329 + 0.01137 i \& 0.18479 + 0.49185 i \& -0.3246 + 0.08248 i \& 0.23979 + 0.35301 i \\
    \ket{\downarrow\uparrow}
    \& {} \& {} \& {} \& {} \& -0.35842 + 0.58475 i \& -0.18602 + 0.36633 i \& 0.54451 - 0.0948 i \& -0.06805 - 0.01823 i \& -0.21426 - 0.06664 i \\
    \ket{\uparrow\downarrow}
    \& {} \& {} \& {} \& {} \& 0.36651 + 0.22964 i \& -0.45339 - 0.20188 i \& 0.05724 + 0.41071 i \& 0.19143 + 0.13307 i \& -0.21499 + 0.5423 i \\
    \ket{\uparrow\uparrow}
    \& {} \& {} \& {} \& {} \& -0.39472 + 0.25686 i \& 0.08382 + 0.07474 i \& -0.31036 - 0.07749 i \& 0.66086 + 0.21797 i \& 0.3691 + 0.2069 i\\
};
\end{tikzpicture}%
}.
\label{eq:gate_qutrits}
\end{equation}
The spin-$\tfrac12$ three-site gate is given by $\hat{U}=\hat{V}\hat{S}_{231 \leftarrow 123}$, where
\begin{equation}
\hat V =
\resizebox{0.96\columnwidth}{!}{%
\begin{tikzpicture}[
    baseline=(M.center),
    blockline/.style={
        draw=black!35,
        dashed,
        line width=0.35pt,
        dash pattern=on 2pt off 1.3pt
    }
]
\matrix (M) [
    matrix of math nodes,
    ampersand replacement=\&,
    left delimiter={(},
    right delimiter={)},
    nodes={
        anchor=center,
        inner xsep=1.7pt,
        inner ysep=1.7pt
    },
    column sep=0pt,
    row sep=0pt
] {
    {}
    \& \bra{001}
    \& \bra{100}
    \& \bra{010}
    \& \bra{101}
    \& \bra{011}
    \& \bra{110}
    \& \bra{000}
    \& \bra{111} \\
    \ket{001}
    \& 0.809016 \& 0.587785 \& {} \& {} \& {} \& {} \& {} \& {} \\
    \ket{100}
    \& -0.587785 \& 0.809017 \& {} \& {} \& {} \& {} \& {} \& {} \\
    \ket{010}
    \& {} \& {} \& -0.113117 - 0.42097 i \& -0.880513 + 0.186239 i \& {} \& {} \& {} \& {} \\
    \ket{101}
    \& {} \& {} \& -0.503511 - 0.745966 i \& 0.373952 - 0.223989 i \& {} \& {} \& {} \& {} \\
    \ket{011}
    \& {} \& {} \& {} \& {} \& 0.02005 - 0.510367 i \& -0.85779 + 0.057628 i \& {} \& {} \\
    \ket{110}
    \& {} \& {} \& {} \& {} \& -0.834526 - 0.206618 i \& 0.069431 - 0.506019 i  \& {} \& {} \\
    \ket{000}
    \& {} \& {} \& {} \& {} \& {} \& {} \& 0.934874 - 0.354978 i \& {} \\
    \ket{111}
    \& {} \& {} \& {} \& {} \& {} \& {} \& {} \& -0.020477  - 0.99979 i \\
};
\end{tikzpicture}%
}.
\label{eq:gate_qubits}
\end{equation}

\SMsection{Currents for Hamiltonian with single charge}

In this section, we show that the equilibrium current vanishes for 1D translationally-invariant Hamiltonians with nearest-neighbor interactions and a single-site conserved charge. This was also proven in Refs.~\onlinecite{PhysRevLett.129.176601,Watanabe2019}; here we prove it using a different method. We also show that this rules out KPZ superdiffusion from the NLFHD mechanism.

Note that this is immediately clear in the case of Hamiltonians with a parity symmetry (i.e., $\hat h_{x,x+1} = \hat h_{x+1, x}$): in that case the currents are parity-antisymmetric and thus their equilibrium expectation values must vanish.

Consider a periodic chain with a
finite-dimensional on-site Hilbert space with the Hamiltonian
\begin{equation}
    \hat H_L=\sum_{x=1}^{L}\hat h_{x,x+1},
    \qquad
    \hat Q_L=\sum_{x=1}^{L}\hat q_x,
    \qquad
    [\hat H_L,\hat Q_L]=0.
    \label{eq:supp-ham-HQ}
\end{equation}
We assume that the total energy $H$ and the total charge \(Q\) are the only conserved quantities of the system for any $L$.

We first show that the two-site Hamiltonian density can be chosen to
conserve the charge on its support.
 Define
\begin{equation}
    \hat k_{0,1}
    := [\hat q_0+\hat q_1,\hat h_{0,1}].
    \label{eq:supp-ham-kdef}
\end{equation}
Eq.~\eqref{eq:supp-ham-HQ} implies that
\(\sum_x \hat k_{x,x+1}=0\).  Expanding \(\hat k\) in a one-site operator
basis consisting of the identity and traceless operators shows that all
genuine two-site components must vanish, while the two one-site
components must have opposite signs.  Hence there is a traceless
one-site operator \(\hat E\) such that
\begin{equation}
    [\hat q_0+\hat q_1,\hat h_{0,1}]
    =\hat E_0-\hat E_1.
    \label{eq:supp-ham-telescope}
\end{equation}
We now remove the telescoping term $\hat{E}$ by a gauge redefinition of the
Hamiltonian density.  Diagonalize the one-site charge,
\begin{equation}
    \hat q:=\sum_{\alpha}q_{\alpha}\hat P_{\alpha},
    \label{eq:supp-ham-qspec}
\end{equation}
where $\alpha$ labels the different eigenvalues and $\hat{P}_\alpha$ is the projector on the eigenspace corresponding to $q_\alpha$. By applying $\hat{P}_\alpha$ on site $0$ from the left and the right and $\hat{P}_\beta$ on site $1$ from the left and the right on \eqref{eq:supp-ham-telescope}, the LHS vanishes and we obtain

\begin{equation}
    \hat P_{\alpha}\hat E\hat P_{\alpha}\otimes\hat P_{\beta}
    =
    \hat P_{\alpha}\otimes
    \hat P_{\beta}\hat E\hat P_{\beta}.
    \label{eq:supp-ham-Ediag}
\end{equation}

Eq.~\eqref{eq:supp-ham-Ediag} can be valid for all \(\alpha,\beta\) only if every diagonal block of \(\hat E\) is equal to the identity multiplied by the same constant for each block.
  
Since \(\hat E\) is traceless, that constant is zero and, therefore,
\begin{equation}
    \hat P_{\alpha}\hat E\hat P_{\alpha}=0
    \qquad \forall\alpha.
    \label{eq:supp-ham-Eoffdiag}
\end{equation}
Thus \(\hat E\) is off-diagonal in the eigenspaces of \(\hat q\). Now we define
\begin{equation}
    \hat a
    :=\sum_{\alpha\ne\alpha'}
      \frac{
      \hat P_{\alpha}\hat E\hat P_{\alpha'}
      }{q_{\alpha}-q_{\alpha'}},
    \qquad
    [\hat q,\hat a]=\hat E.
    \label{eq:supp-ham-adef}
\end{equation}
Using Eq.~\eqref{eq:supp-ham-telescope}, one immediately finds that
\begin{equation}
    \hat h'_{0,1}
    :=\hat h_{0,1}-(\hat a_0-\hat a_1)
    \qquad\Longrightarrow\qquad
    [\hat h'_{0,1},\hat q_0+\hat q_1]=0.
    \label{eq:supp-ham-gdef}
\end{equation}
Therefore, the general form of the Hamiltonian is 
\begin{equation}
    \hat h_{0,1}=\hat a_0-\hat a_1+\hat h'_{0,1},
    \qquad
    [\hat h'_{0,1},\hat q_0+\hat q_1]=0.
    \label{eq:supp-ham-general-h}
\end{equation}
Notice that the $\hat{a}$ terms do not change the
periodic Hamiltonian because
\(\sum_x(\hat a_x-\hat a_{x+1})=0\).  On an open chain, they change the
Hamiltonian only by boundary terms, which do not affect the bulk equilibrium
currents in the thermodynamic limit.  We henceforth work in this gauge, so the general Hamiltonian commuting with a charge satisfies $[\hat{h}_{01},\hat{q}_0+\hat{q}_1]=0$.

The charge and energy currents are
\begin{align}
    \hat \jmath^{\,q}_{x,x+1}
    &=\mathrm{i}[\hat h_{x,x+1},\hat q_{x+1}],
    \label{eq:supp-ham-jq}\\
    \hat \jmath^{\,e}_{x}
    &=\mathrm{i}[\hat h_{x-1,x},\hat h_{x,x+1}],
    \label{eq:supp-ham-je}
\end{align}
obeying the continuity equations
\begin{align}
    \frac{\mathrm{d}\hat q_x}{\mathrm{d}t}
    &=\hat \jmath^{\,q}_{x-1,x}
      -\hat \jmath^{\,q}_{x,x+1},
    \label{eq:supp-ham-cont-q}\\
    \frac{\mathrm{d}\hat h_{x,x+1}}{\mathrm{d}t}
    &=\hat \jmath^{\,e}_{x}
      -\hat \jmath^{\,e}_{x+1}.
    \label{eq:supp-ham-cont-e}
\end{align}
Now consider an open chain and the two-parameter equilibrium state
\begin{equation}
    \hat\rho_L(\beta,\mu)
    =\frac{1}{Z_L(\beta,\mu)}
      \exp[-\beta\hat H_L+\mu\hat Q_L],
    \label{eq:supp-ham-rho}
\end{equation}
where \(\mu\) is the dimensionless chemical potential. Bondwise charge
conservation ensures \([\hat H_L,\hat Q_L]=0\) also for the open chain,
and hence \([\hat\rho_L,\hat H_L]=0\), i.e., the state is stationary. At the
left boundary, Eqs.~\eqref{eq:supp-ham-cont-q} and
\eqref{eq:supp-ham-cont-e} consequently give
\begin{equation}
    0=\left\langle\dot{\hat q}_1\right\rangle_{\beta,\mu}
      =-\left\langle\hat \jmath^{\,q}_{1,2}\right\rangle_{\beta,\mu},
    \qquad
    0=\left\langle\dot{\hat h}_{1,2}\right\rangle_{\beta,\mu}
      =-\left\langle\hat \jmath^{\,e}_{2}\right\rangle_{\beta,\mu}.
    \label{eq:supp-ham-boundary-current}
\end{equation}
Stationarity of the remaining local densities then propagates these
equalities through the chain.  Therefore,
\begin{equation}
    \left\langle\hat \jmath^{\,q}_{x,x+1}\right\rangle_{\beta,\mu}=0,
    \qquad
    \left\langle\hat \jmath^{\,e}_{x}\right\rangle_{\beta,\mu}=0
    \label{eq:supp-ham-zero-current-open}
\end{equation}
on every bond of the open chain. Taking the bulk thermodynamic limit gives
the equilibrium Euler-current functions
\begin{equation}
    \jmath_q(\beta,\mu)=0,
    \qquad
    \jmath_e(\beta,\mu)=0.
    \label{eq:supp-ham-zero-current}
\end{equation}
Importantly, the conclusion is not based on time-reversal or inversion symmetry;
it follows from Hamiltonian stationarity, locality, and the absence of an
additional conserved quantity. Note that the argument can easily be extended to Hamiltonians coupling any finite number of neighboring sites by grouping those sites into supersites, which are coupled again by a nearest-neighbor Hamiltonian. It can also be straightforwardly extended to the case of multiple conserved quantities, as long as they are also sums of single-site terms. The only option for having non-vanishing current expectation values is to have at least one further conserved quantity that couples two or more sites~\cite{PhysRevResearch.4.013043}. This happens, for instance, in integrable systems, where the current expectation values can be computed using the framework of generalized hydrodynamics~\cite{doyon2020lecture}.

In particular, the quadratic
self-coupling that would generate the KPZ mechanism is
absent for any Gibbs state. A Floquet circuit is constrained only
after a complete period: its individual gates or layers need not
conserve the charge on their support. Furthermore, our circuits do not
admit a charge-conserving version with open boundary conditions. Instead, a spatially
uniform current may therefore circulate in a stationary state on a
ring. This is how the circuits considered in the main text evade the
Hamiltonian obstruction above.

\SMsection{Two-site spin-$\frac12$ circuits and equilibrium currents}

In this section, we will explain that all two-site brickwork spin-$\frac12$ circuits with a conserved quantity that is a sum of single site terms are either integrable or have zero equilibrium current. Throughout this section, by integrability we mean the existence, in the thermodynamic limit, of infinitely many linearly independent, mutually commuting extensive conserved quantities with finite-range densities. We do not require these conserved quantities to arise from a Yang–Baxter construction.

The time evolution consists of repeated applications of two layers
\begin{equation} \hat{\mathbb U} := \hat{\mathbb U}_{\mathrm{o}}\hat{\mathbb U}_{\mathrm{e}}, ~~~~~ \hat{\mathbb U}_{\mathrm{o}} := \prod_{n=0}^{L-1} \hat U_{2n+1,2n+2}, \qquad \hat{\mathbb U}_{\mathrm{e}} := \prod_{n=1}^{L} \hat U_{2n,2n+1}.
\end{equation}
The conserved charge of range 1 is given by 
\begin{align}
    \hat{Q} &= \sum_n \hat{q}\upd{o}_{2n-1} + \hat{q}\upd{e}_{2n},~~~~~~~[\hat{Q},\hat{\mathbb{U}}]=0,\label{equ:sm_2qubit_charge}
\end{align}
Subtracting the identity component, we take both Hermitian densities to be
traceless. Their spectra are given by
$\operatorname{spec}(\hat q^a)=\{\alpha_a/2,-\alpha_a/2\}$,
where $\alpha_a\geq0$ and $\alpha_o+\alpha_e>0$.

The full period conservation implies that
\begin{equation}
    \hat{\mathbb{U}}_\mathrm{o}^{\dagger}\hat{Q}\hat{\mathbb{U}}_\mathrm{o}=\hat{\mathbb{U}}_\mathrm{e} \hat{Q} \hat{\mathbb{U}}_\mathrm{e}^{\dagger} =: \hat C,\label{eq:sm_inter_C}
\end{equation}
where $\hat C$ is the charge propagated for half a period.
We need to deduce the most general condition for this equality to hold. Let's define 
\begin{align}
\begin{aligned}
\hat A&:=\hat U^\dagger
(\hat{q}\upd{o}\otimes \hat{I}+\hat{I}\otimes\hat q\upd{e}) \hat{U},\
~~~\hat B:=\hat{U}
(\hat{q}\upd{e}\otimes \hat{I}+\hat{I}\otimes\hat{q}\upd{o})\hat U^\dagger.
\end{aligned}
\end{align}
Now Eq. \eqref{eq:sm_inter_C} can be rewritten as
\begin{equation}
\hat{C}
=\sum_n\hat A_{2n-1,2n}
=\sum_n\hat B_{2n,2n+1}.\label{eq:sm:ABdef}
\end{equation}
Now let us expand the Hermitian traceless 2-site operator $\hat{A}$ in the Pauli basis
\begin{equation}
\hat A=\hat c^o\otimes \hat{I}+\hat{I}\otimes\hat c^e
+\sum_{\mu,\nu=x,y,z}d_{\mu\nu}
\hat{\sigma}^\mu\otimes\hat{\sigma}^\nu,
\end{equation}
where 
\begin{equation}
\hat c\upd{o}:=\tfrac12\operatorname{tr}_2\hat A,
\qquad
\hat c\upd{e}:=\tfrac12\operatorname{tr}_1\hat A,~~~~~~~d_{\mu\nu}:=\tfrac14 \operatorname{tr}\left(\hat{\sigma}^\mu \otimes \hat{\sigma}^\nu ~ \hat{A}\right).
\end{equation}
Notice that from orthogonality of the Pauli strings and Eq. \eqref{eq:sm:ABdef} it follows that
\begin{align}
\begin{aligned}
d_{\mu\nu}
&=2^{-2L}\operatorname{tr}
\left(\hat{\sigma}^\mu_{2n-1}\hat{\sigma}^\nu_{2n}\hat C\right)\
=2^{-2L}\sum_m\operatorname{tr}
\left(\hat{\sigma}^\mu_{2n-1}\hat{\sigma}^\nu_{2n}
\hat B_{2m,2m+1}\right)
=0.
\end{aligned}
\end{align}
Thus $\hat{C}$ contains no two-site terms and can be written as
\begin{equation}
    \hat C=\sum_n(\hat c^\mathrm{o}_{2n-1}+\hat c^\mathrm{e}_{2n}),
\label{eq:qubit_current_intermediate_local}
\end{equation}
with traceless one-site densities. The action of the gate on one-site densities is given by
\begin{align}
    \hat U^\dagger
    (\hat q^\mathrm{o}\otimes \hat{I}+\hat{I}\otimes\hat q^\mathrm{e})\hat U
    &=\hat c^\mathrm{o}\otimes \hat{I}+\hat{I}\otimes\hat c^\mathrm{e},
    \label{eq:qubit_current_local_odd}\\
    \hat U^\dagger
    (\hat c^\mathrm{e}\otimes \hat{I}+\hat{I}\otimes\hat c^\mathrm{o})\hat U
    &=\hat q^\mathrm{e}\otimes \hat{I}+\hat{I}\otimes\hat q^\mathrm{o}.
    \label{eq:qubit_current_local_even}
\end{align}

Let $\operatorname{spec}(\hat c^a)=\{\gamma_a/2,-\gamma_a/2\}$,
with $\gamma_a\geq0$. Notice that a unitary rotation of the charge density in Eq.~\eqref{eq:qubit_current_local_odd}  does not change the spectrum, so $\text{spec}(\hat q^\mathrm{o}\otimes \hat{I}+\hat{I}\otimes\hat q^\mathrm{e})=\text{spec}(\hat c^\mathrm{o}\otimes \hat{I}+\hat{I}\otimes\hat c^\mathrm{e})$. That implies equality of the four
eigenvalues $\pm(\alpha_o+\alpha_e)/2$ and
$\pm(\alpha_o-\alpha_e)/2$ with $\pm(\gamma_o+\gamma_e)/2$ and $\pm(\gamma_o-\gamma_e)/2$. Hence we have two options
\begin{equation}
    \gamma_\mathrm{o}=\alpha_\mathrm{o},~~\gamma_\mathrm{e}=\alpha_\mathrm{e} ~~~~\text{or}~~~~\gamma_\mathrm{o}=\alpha_\mathrm{e},~~\gamma_\mathrm{e}=\alpha_\mathrm{o}.
    \label{eq:qubit_current_spectra}
\end{equation}
Since $\hat C$ is the charge propagated for half a period, option 1
implies that the charge is conserved (at the level of the spectrum) also after
half a period, whereas option 2 implies that the charge is ``translated'' after
half a period. Our spin-$1$ construction from the main text corresponded to
option 2, as is evident
from Eq.~\eqref{eq:q_drift}.
We show below that for spin-$\tfrac{1}{2}$
\begin{align}
    1)~\text{if}~(\gamma_{\mathrm{o}},\gamma_{\mathrm{e}})=(\alpha_{\mathrm{o}},\alpha_{\mathrm{e}}) ~&\Rightarrow ~~j(\mu)=\text{const}\\ 
    2) ~\text{if}~(\gamma_{\mathrm{o}},\gamma_{\mathrm{e}})=(\alpha_{\mathrm{e}},\alpha_{\mathrm{o}})~~\text{and} ~~\alpha_{\mathrm{e}} \neq \alpha_{\mathrm{o}}~&\Rightarrow ~~\text{the circuit is integrable}
\end{align}

\textbf{1).} Let's find the current in the most general case of conservation of a one-site charge given by Eqs. \eqref{eq:qubit_current_local_odd}, \eqref{eq:qubit_current_local_even}. Using the charge density $\hat{q}(x)=\hat{q}_{2x-1}^\mathrm{o}+\hat{q}_{2x}^\mathrm{e}$, the continuity equation which determines the current reads
\begin{equation}
    \hat{\mathbb{U}}^{\dagger} \hat{q}(x) \hat{\mathbb{U}}-\hat{q}(x)=\hat{\jmath}(x-\tfrac12)-\hat{\jmath} (x+\tfrac12).
\end{equation}
Repeating the derivation of the current from the main text, it can be shown that the continuity equation is satisfied by the following current:
\begin{equation}
    \hat{\jmath}(x+\tfrac12)=\hat{q}^\mathrm{e}_{2x}-\hat{U}^\dagger_{2x,2x+1} \hat{c}^\mathrm{e}_{2x}\hat{U}_{2x,2x+1}.
\end{equation}
Similarly to the case in the main text, the equilibrium expectation value of the current factorizes as
\begin{equation}
    \jmath(\mu)=\frac{\text{tr}(\hat{\jmath}(x+\tfrac12) e^{\mu \hat{Q}})}{\text{tr}( e^{\mu \hat{Q}})}=\frac{\text{tr}(\hat{q}^\mathrm{e} e^{\mu \hat{q}^\mathrm{e}})}{\text{tr}( e^{\mu \hat{q}^\mathrm{e}})}-\frac{\text{tr}(\hat{c}^\mathrm{e} e^{\mu \hat{c}^\mathrm{e}})}{\text{tr}( e^{\mu \hat{c}^\mathrm{e}})}=\frac{\alpha_\mathrm{e}}{2}\tanh \frac{\mu \alpha_\mathrm{e}}{2}-\frac{\gamma_\mathrm{e}}{2}\tanh \frac{\mu \gamma_\mathrm{e}}{2}.
    \label{eq:sm_current_spin12}
\end{equation}
This implies that for the case of $(\gamma_{\mathrm{o}},\gamma_{\mathrm{e}})=(\alpha_{\mathrm{o}},\alpha_{\mathrm{e}})$, the equilibrium current vanishes $j(\mu)=0$. Notice that the case of $\alpha_{\mathrm{o}}=\alpha_{\mathrm{e}}$ is also covered in this case.

\textbf{2).}  Here we will show that in the case of $(\gamma_\mathrm{o},\gamma_\mathrm{e})=(\alpha_\mathrm{e},\alpha_\mathrm{o})$ with $\alpha_\mathrm{o} \neq \alpha_\mathrm{e}$, we can construct an infinite number of conserved quantities in the thermodynamic limit. The conserved quantities are gliders –- under the time evolution, their density does not spread, but glides along lightcone, similar to the charges found in dual-unitary circuits~\cite{Gliders_DU,Gliders_Masanes}.

First notice that $(\hat{q}^\mathrm{o})^2=\alpha_\mathrm{o}^2 \frac{\hat{I}}{4}$ and $(\hat{q}^\mathrm{e})^2=\alpha_\mathrm{e}^2 \frac{\hat{I}}{4}$. That allows us to simplify the powers of the density $\hat{q}=\hat{q}^\mathrm{o} \otimes \hat{I}+\hat{I}\otimes \hat{q}^\mathrm{e}$, particularly
\begin{equation}
   \hat{q}^\mathrm{o} \otimes \hat{I}=\frac{(3\alpha_\mathrm{o}^2+\alpha_\mathrm{e}^2)\hat{q}-4 \hat{q}^3}{2(\alpha_\mathrm{o}^2-\alpha_\mathrm{e}^2)}.
\end{equation}

Since unitary conjugation preserves polynomial relations, Eq. \eqref{eq:qubit_current_local_odd} allows us to replace $\hat{q}$ in this polynomial by $\hat c^{\mathrm o}\otimes \hat{I}+\hat{I}\otimes\hat c^{\mathrm e}$. In the case under consideration, the spectral magnitudes are exchanged $
(\hat c^{\mathrm o})^2=\frac{\alpha_{\mathrm e}^2}{4}\hat{I},~
(\hat c^{\mathrm e})^2=\frac{\alpha_{\mathrm o}^2}{4}\hat{I}$.
The same polynomial identity therefore isolates the contribution on the second site, giving
\begin{equation}
    \hat{U}^{\dagger} (\hat{q}^\mathrm{o}\otimes \hat{I})\hat{U}=\hat{I} \otimes \hat{c}^\mathrm{e},~~~~\hat{U}^{\dagger} (\hat{I}\otimes\hat{q}^\mathrm{e} )\hat{U}= \hat{c}^\mathrm{o}\otimes \hat{I}.
\end{equation}
Here the second equality follows by subtracting the first from Eq. \eqref{eq:qubit_current_local_odd}. Thus each one-site density is transferred to the opposite site without developing any two-site components.

Applying the same argument to Eq.~\eqref{eq:qubit_current_local_even}, the next layer transfers these intermediate densities to the neighboring sites and restores their original form. Over a full period, the odd- and even-sublattice densities are therefore translated by two sites in opposite directions:
\begin{equation}
    \hat{\mathbb{U}}^\dagger \hat{q}^\mathrm{o}_{2n-1}\hat{\mathbb{U}}=\hat{q}^\mathrm{o}_{2n+1}~,~~~~\hat{\mathbb{U}}^\dagger \hat{q}^\mathrm{e}_{2n}\hat{\mathbb{U}}=\hat{q}^\mathrm{e}_{2n-2}~.
\end{equation}
In other words, the $\hat{q}^\mathrm{o}$ and $\hat{q}^\mathrm{e}$ are exact gliders, moving with velocity $\pm 2$. Furthermore, products of densities on both sublattices yield
infinitely many extensive conserved quantities, e.g.,
\begin{equation}
    \hat{Q}_r^\mathrm{o}=\sum_n \hat{q}^\mathrm{o}_{2n-1}\hat{q}^\mathrm{o}_{2n+2r-1}~~,~~~~
    \hat{Q}_r^\mathrm{e}=\sum_n \hat{q}^\mathrm{e}_{2n} \hat{q}^\mathrm{e}_{2n+2r}.
\end{equation}
These operators commute with $\hat{\mathbb U}$ and with one another because they
act nontrivially on different sublattices. Each has a density supported on
$2r+1$ consecutive sites, and distinct separations give linearly independent
charges on sufficiently large rings. Therefore, the circuit is integrable in the
sense above.

\SMsection{Conserved quantities and current for three-site qubit gates}

Three-site qubit circuits can be studied using a construction analogous to that for two-site gates. The decomposition $\hat{\mathbb U}
=(\hat{\mathbb U}_2\hat T^\dagger)^3\hat T^3$ implies that, for a charge invariant under translation by three sites, $[\hat Q,\hat T^3]=0,
~~
[\hat Q,\hat{\mathbb U}_2\hat T^\dagger]=0
~\Longrightarrow~
[\hat Q,\hat{\mathbb U}]=0.$ We can, therefore, construct conserved charges by imposing conservation under $\hat{\mathbb U}_2\hat T^\dagger$, which is sufficient for conservation under the full-period evolution. The condition $[\hat{Q},\hat{\mathbb U}_2\hat T^\dagger]=0$ can be pictorially represented
\begin{align}
   \picR[scale=0.027][+0.0em]{Ud3qU3}\label{equ:sm_3qubit_sum} \;\; ,
\end{align}
where 
    $\hat{q} = \hat{q}^2 \otimes \hat{I} \otimes \hat{I} + \hat{I} \otimes \hat{q}^0 \otimes  \hat{I} + \hat{I} \otimes \hat{I} \otimes \hat{q}^1$, or pictorially
\begin{align}
\picR[scale=0.03][-0.1em]{q3_def_201} \; ,\label{equ:sm_3site_q_def}
\end{align}
with $\hat{q}^0, \hat{q}^1$ and $\hat{q}^2$ being the charges on the $0$th, $1$st and $2$nd sites. The definition of the supercell charge density $\hat{q}$  in \eqref{equ:sm_3site_q_def} turns out to be a mathematically convenient choice.

From \eqref{equ:sm_3qubit_sum}, we find the condition
\begin{align}
\picR[scale=0.027][-0.2em]{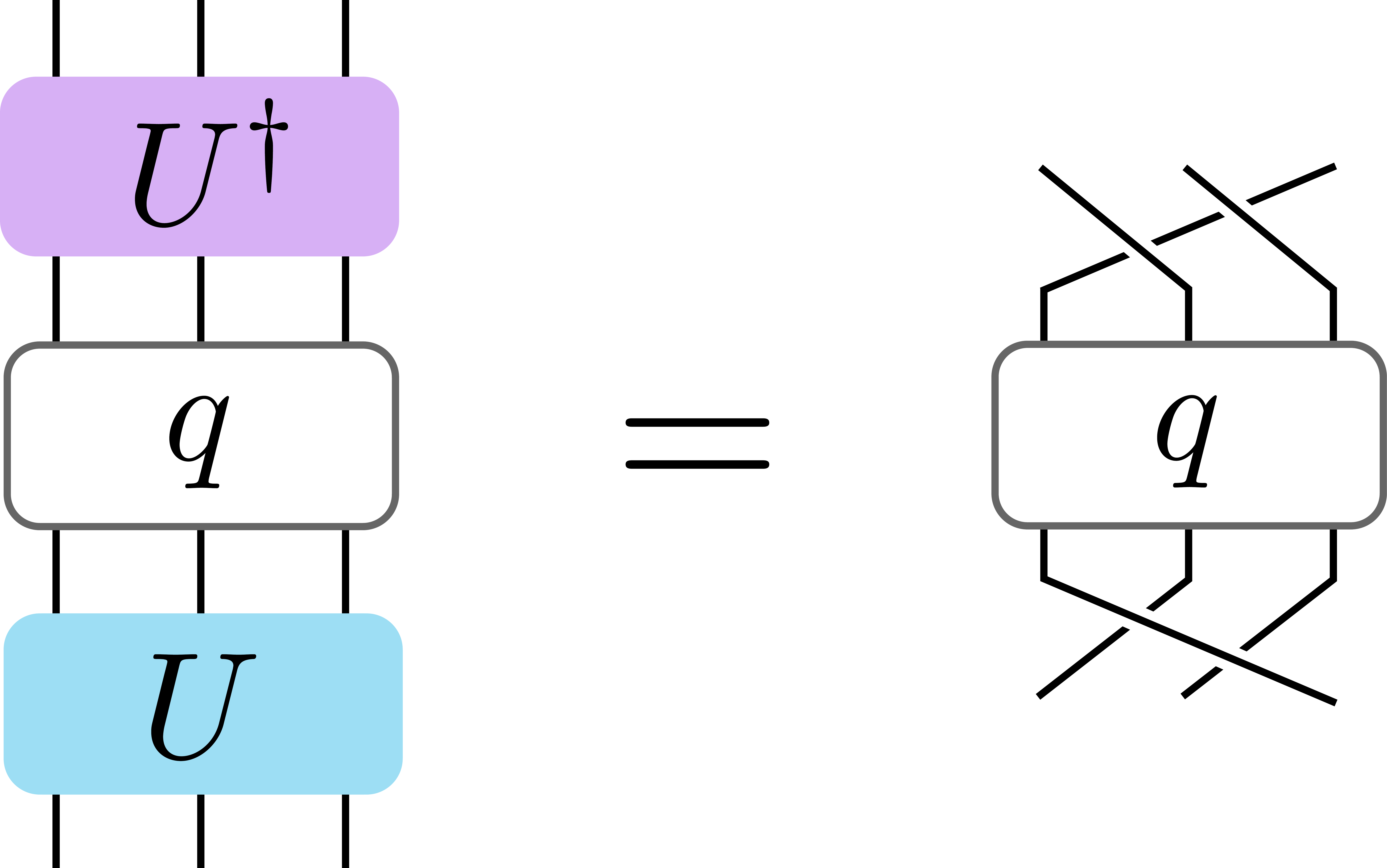} \; .
\end{align}
Since $\hat{S}_{312 \leftarrow 123} = (\hat{S}_{231 \leftarrow 123})^\dag = (\hat{S}_{231 \leftarrow 123})^{-1}$, this implies that $\hat{V} = \hat{U} \hat{S}_{312 \leftarrow 123} = \hat{U}\sum_{a,b,c} \ketbra{bca}{abc}$ commutes with $\hat{q}$, i.e.,
\begin{align}
    \qty[\hat{V},\hat{q}] &= 0.
\end{align}
Therefore, $\hat V$ can be chosen independently within each
eigenspace of $\hat q$, giving the block structure of the gate.

We next derive the corresponding current. Applying the local
conservation identity successively through the three layers
gives the discrete continuity equation
\begin{align}
  \picR[width=0.92\columnwidth][-0.1em]{q3_continuity}
\end{align}
with the current operator given by
\begin{align}
\picR[width=0.84\columnwidth][-0.2em]{j3def}\label{eq:sm_3qubit_current_op}
\end{align}
where
\begin{align}
\picR[width=0.27\columnwidth][-0.1em]{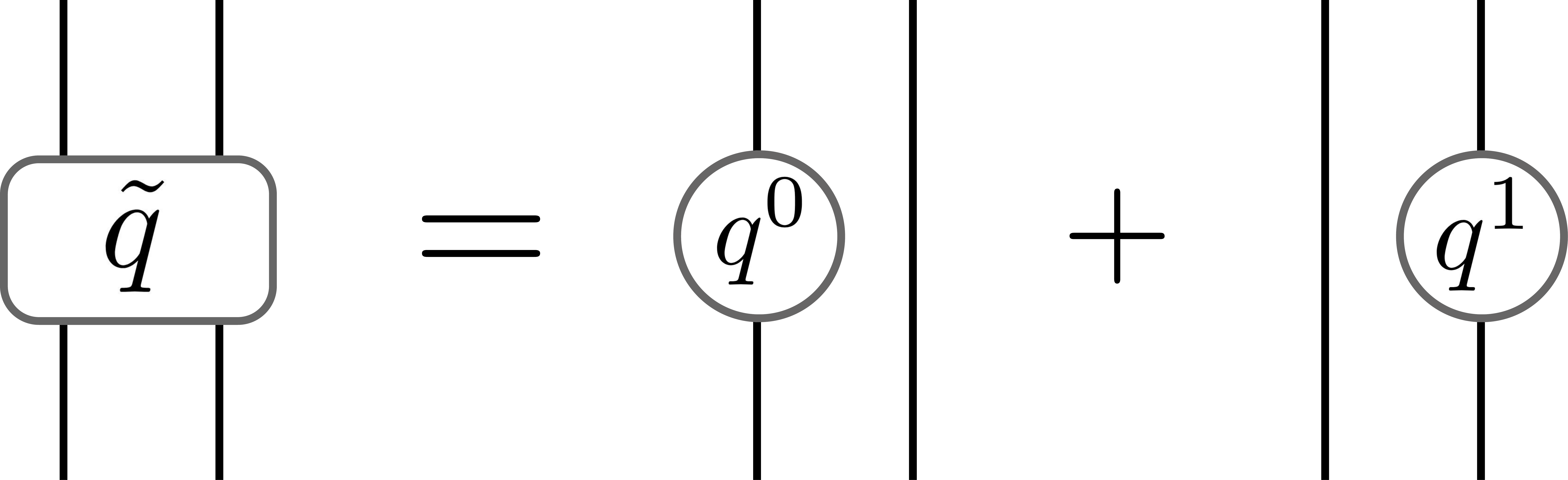} \; .
\end{align}
We now evaluate the current in the equilibrium state
$\hat\rho_\mu=Z(\mu)^{-1}e^{\mu\hat Q}$.
Since $\hat Q$ is a sum of one-site operators, this state
factorizes:
\begin{equation}
    \hat\rho_\mu
    =\bigotimes_n
    \left(
        \hat\rho_\mu^0\otimes
        \hat\rho_\mu^1\otimes
        \hat\rho_\mu^2
    \right),
    \qquad
    \hat\rho_\mu^a
    =\frac{e^{\mu\hat q^a}}
    {\operatorname{tr}e^{\mu\hat q^a}}.
\end{equation}
Although this state is stationary under a complete period,
individual layers generally change its sublattice structure.
Indeed, $\hat U=\hat V\hat S$ and $[\hat V,\hat q]=0$ imply
\begin{equation}
    \hat U
    \left(
        \hat\rho_\mu^0\otimes
        \hat\rho_\mu^1\otimes
        \hat\rho_\mu^2
    \right)
    \hat U^\dagger
    =
    \hat\rho_\mu^2\otimes
    \hat\rho_\mu^0\otimes
    \hat\rho_\mu^1.
\end{equation}
To see this, first apply $\hat S$, which permutes the three
factors. The resulting product is proportional to
$e^{\mu\hat q}$ and therefore commutes with $\hat V$.
Consequently, the sublattice pattern evolves as
$(0,1,2)\to(2,0,1)\to(1,2,0)\to(0,1,2)$ under the three
successive layers.

Using the cyclicity of the trace, the gates in the two positive
terms of the current operator can be transferred onto the
Gibbs state. The intermediate sublattice patterns above
then show that the first two terms on the RHS of Eq.~\eqref{eq:sm_3qubit_current_op} each have the expectation value
$\expval{\hat{q}^0}_\mu+\expval{\hat{q}^1}_\mu$. The remaining two terms contribute
$-\expval{\hat{q}}_\mu$ and $-\expval{\hat{q}^2}_\mu$, respectively. We therefore obtain
\begin{align}
    \expval{\hat{q}}_\mu &= \expval{\hat{q}^0}_\mu + \expval{\hat{q}^1}_\mu + \expval{\hat{q}^2}_\mu\\
    \expval{\hat{\jmath}}_\mu &= \expval{\hat{q}^0}_\mu + \expval{\hat{q}^1}_\mu -2  \expval{\hat{q}^2}_\mu.
\end{align}
Similar to the qutrit case, the equilibrium current depends only on the one-site
charge spectra and the circuit geometry. It is independent
of the unitary matrices chosen within the charge blocks.

Note that the $\hat{q}^a$ can always be made traceless (by subtracting their trace) and diagonal (by changing the local basis). In the qubit case, we can therefore write without loss of generality 
\begin{equation}
   \hat q^a=2\alpha_a \hat{S}^z. 
   \label{eq:sm_charge_3qubit}
\end{equation}
Thus the equilibrium expectation values are
\begin{align}
q(\mu)&=\sum_{a=0}^2\alpha_a\tanh(\mu\alpha_a),\\
    j(\mu)
    &=\alpha_0\tanh(\mu\alpha_0)
    +\alpha_1\tanh(\mu\alpha_1)
    -2\alpha_2\tanh(\mu\alpha_2).
    \label{eq:sm_equil}
\end{align}
If $\alpha_a$ are generic numbers, then $\hat{q}$ will have no degeneracies and only diagonal $\hat{V}$ are possible, implying that $\hat{U}$ will be a super-integrable SWAP gate dressed by phases. We need to ensure that the qubit circuit we consider has only one conserved quantity.

Therefore, only very restricted options for $\alpha_i$ exist. We list all of
them in Table~\ref{tbl:all_3site}. Furthermore, we checked the number of
conserved charges at small support $l$ using the so-called truncated
propagator~\cite{prosenRuelleResonancesQuantum2002}. In short, the truncated
propagator is the Heisenberg propagator of observables $\mathcal{U}[A] :=
\mathbb{U}^\dagger A \mathbb{U}$ projected to the space of local observables
with densities supported on a finite number of sites $l$, i.e., to the space
observables of the form $A = \sum_j \hat T^{-rj} a \hat T^{rj}$, where $a$ acts
non-trivially on an interval of at most $l$ sites. The obtained truncated
propagator $\mathcal U_l$ is a non-unitary matrix with each charge corresponding
to the eigenvalue $1$. Matrix elements of the truncated propagator in the
thermodynamic limit $L \to \infty$ can be obtained exactly and its finite $l$
version numerically diagonalized. The number of independent charges with support
on at most $l$ is then equal to the degeneracy of the eigenvalue $1$ at some
support $l$. For implementation details in the considered systems see
Ref.~\onlinecite{urban3site}.

\begin{table*}[h!]
\centering
\small
\setlength{\tabcolsep}{4pt}

\begin{minipage}[t]{0.45\textwidth}
\centering
\begin{tabular*}{\linewidth}{@{\extracolsep{\fill}}c|ccc|*{8}{r}@{}}
\hline
 & & & & \multicolumn{8}{c@{}}{\shortstack{Number of charges\\ with support up to $l=$}}\\
Index & $\alpha_0$ & $\alpha_1$ & $\alpha_2$ & 1 & 2 & 3 & 4 & 5 & 6 & 7 & 8\\
\hline
0  & $0$   & $0$   & $1$   & 1 & 1 & 1 & 2 & 3 & 3 & 5 & 8\\
1  & $0$   & $1/2$ & $1$   & 2 & 2 & 2 & 4 & 8 & 8 & * & *\\
2  & $0$   & $1$   & $0$   & 1 & 1 & 1 & 2 & 2 & 2 & 4 & 4\\
3  & $0$   & $1$   & $1/2$ & 2 & 2 & 2 & 4 & 8 & 8 & * & *\\
\textbf{4} & $\mathbf{0}$ & $\mathbf{1}$ & $\mathbf{1}$
   & \textbf{1} & \textbf{1} & \textbf{1} & \textbf{1} & \textbf{1} & \textbf{1} & \textbf{1} & \textbf{1}\\
5  & $1/4$ & $1/4$ & $1$   & 2 & 3 & 3 & 4 & * & * & * & *\\
6  & $1/4$ & $3/4$ & $1$   & 2 & 3 & 3 & 4 & * & * & * & *\\
7  & $1/4$ & $1$   & $1/4$ & 2 & 2 & 2 & 3 & 3 & 3 & 5 & 5\\
8  & $1/4$ & $1$   & $3/4$ & 2 & 3 & 3 & 4 & * & * & * & *\\
9  & $1/2$ & $0$   & $1$   & 2 & 2 & 2 & 4 & 8 & 8 & * & *\\
10 & $1/2$ & $1/2$ & $1$   & 1 & 2 & 2 & 2 & 4 & 4 & 4 & *\\
11 & $1/2$ & $1$   & $0$   & 2 & 3 & 4 & 9 & * & * & * & *\\
\textbf{12} & $\mathbf{1/2}$ & $\mathbf{1}$ & $\mathbf{1/2}$
   & \textbf{1} & \textbf{1} & \textbf{1} & \textbf{1} & \textbf{1} & \textbf{1} & \textbf{1} & \textbf{1}\\
13 & $1/2$ & $1$   & $1$   & 2 & 2 & 2 & 3 & 3 & 3 & 5 & 5\\
\hline
\end{tabular*}\end{minipage}\hfill%
\begin{minipage}[t]{0.47\textwidth}
\centering
\begin{tabular*}{\linewidth}{@{\extracolsep{\fill}}c|ccc|*{8}{r}@{}}
\hline
 & & & & \multicolumn{8}{c@{}}{\shortstack{Number of charges\\ with support up to $l=$}}\\
Index & $\alpha_0$ & $\alpha_1$ & $\alpha_2$ & 1 & 2 & 3 & 4 & 5 & 6 & 7 & 8\\
\hline
14 & $3/4$ & $1/4$ & $1$   & 2 & 3 & 3 & 4 & $*$ & $*$ & $*$ & $*$\\
15 & $3/4$ & $1$   & $1/4$ & 2 & 3 & 3 & 4 & $*$ & $*$ &  $*$   &  $*$\\
16 & $1$   & $0$   & $0$   & 1 & 1 & 1 & 2 & 2 & 2 & 4 & 4\\
17 & $1$   & $0$   & $1/2$ & 2 & 2 & 2 & 4 & 8 & 8 & $*$ & $*$\\
\textbf{18} & $\mathbf{1}$ & $\mathbf{0}$ & $\mathbf{1}$
   & \textbf{1} & \textbf{1} & \textbf{1} & \textbf{1} & \textbf{1} & \textbf{1} & \textbf{1} & \textbf{1}\\
19 & $1$   & $1/4$ & $1/4$ & 2 & 2 & 2 & 3 & 3 & 3 & 5 & 5\\
20 & $1$   & $1/4$ & $3/4$ & 2 & 3 & 3 & 4 & $*$ & $*$ & $*$ & $*$\\
21 & $1$   & $1/2$ & $0$   & 2 & 3 & 4 & 9 & $*$ & $*$ & $*$ & $*$\\
\textbf{22} & $\mathbf{1}$ & $\mathbf{1/2}$ & $\mathbf{1/2}$
   & \textbf{1} & \textbf{1} & \textbf{1} & \textbf{1} & \textbf{1} & \textbf{1} & \textbf{1} & \textbf{1}\\
23 & $1$   & $1/2$ & $1$   & 2 & 2 & 2 & 3 & 3 & 3 & 5 & 5\\
24 & $1$   & $3/4$ & $1/4$ & 2 & 3 & 3 & 4 & $*$ & $*$ & $*$ & $*$\\
25 & $1$   & $1$   & $0$   & 1 & 2 & 2 & 2 & 6 & 6 & 6 & $*$\\
26 & $1$   & $1$   & $1/2$ & 2 & 3 & 3 & 4 & $*$ & $*$ & $*$ & $*$\\
\textbf{27} & $\mathbf{1}$ & $\mathbf{1}$ & $\mathbf{1}$
   & \textbf{1} & \textbf{1} & \textbf{1} & \textbf{1} & \textbf{1} & \textbf{1} & \textbf{1} & \textbf{1}\\
\hline
\end{tabular*}
\end{minipage}

\caption{Number of charges in circuits conserving the
charge~\eqref{eq:sm_3qubit_current_op} as obtained by the truncated propagator method.
Tested for 10 different random representatives. Where the number of
charges varies between the representatives, we write $*$.}
\label{tbl:all_3site}
\end{table*}

Systems
that generically conserve only $\hat Q$ are marked in bold; they have indices $\textbf{4}, \textbf{12},
\textbf{18}, \textbf{22} ~\text{and} ~\textbf{27}$. For gates with indices $\textbf{4}, ~\textbf{18}$ and $\textbf{27}$ our hydrodynamic theory predicts vanishing
superdiffusion constant $\lambda_B = 0$ and thus diffusive transport. The gate with
index $\textbf{22}$ is the one considered in the main text.
The gate with index $\textbf{12}$ has the same $\jmath(q)$ as gate with index $\textbf{22}$ and thus the same
NLFHD predictions.

For the gate considered in the main text,
$(\alpha_0,\alpha_1,\alpha_2)=(1,\tfrac12,\tfrac12)$,
equilibrium expectation values \eqref{eq:sm_equil} reduce to
\begin{equation}
    q(\mu)=\tanh\mu+\tanh\frac{\mu}{2},
    \qquad
    j(\mu)=\tanh\mu-\frac12\tanh\frac{\mu}{2},
\end{equation}
recovering the expressions in the main text.

\SMsection{Quantum chaos of quantum circuits}
Indications that a quantum circuit is chaotic can be obtained by studying its
level-spacing distribution~\cite{Haake}. Namely, diagonalizing the propagator
$\hat{\mathbb{U}} \ket{\phi_j} = \mathrm{e}^{\mathrm{i} \phi_j} \ket{\phi_j}$
and sorting $\phi_j$ by increasing magnitude, the level spacing is defined as
$s_j := (\phi_{j + 1} - \phi_{j})/(2 \pi \mathcal{N})$, where $\mathcal N$ is
the Hilbert space dimension. For a chaotic system, the distribution of $s_j$ is
expected to conform to the predictions of one of the circular ensembles of
random matrix theory (RMT) in every irreducible symmetry subsector, while for
generic integrable systems it is expect to follow the predictions of the Poissonian
ensemble. In contrast to Hamiltonian systems, the density of states in
generic Floquet systems is approximately uniform, so no unfolding~\cite{Haake}
is necessary.

\begin{figure}[hb]
    \centering
    \includegraphics[width=0.47\textwidth]{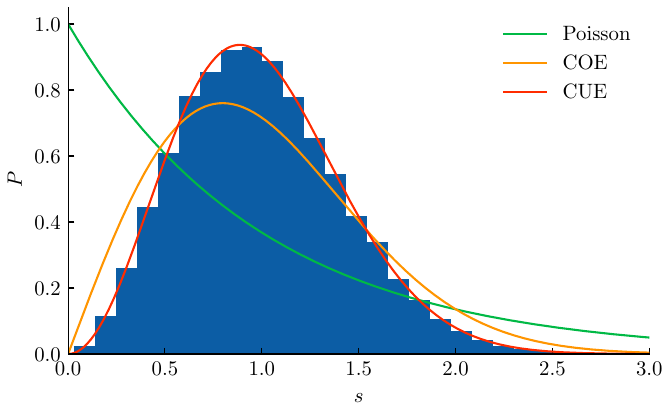}
    \includegraphics[width=0.47\textwidth]{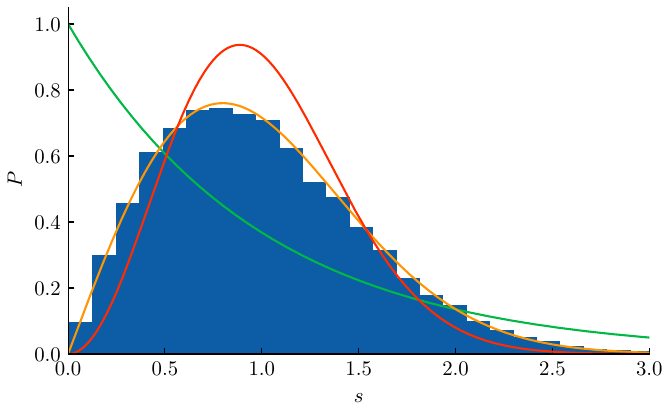}
    \caption{Momentum, charge and space-time symmetry resolved level spacing
    distribution in the spin-1 \eqref{eq:gate_qubits} case with $L = 10$ sites
    (left) and in the spin-$\tfrac12$ \eqref{eq:gate_qutrits} case with $L =
    15$ (right). The level-spacing is calculated in a finite system with
    periodic boundary conditions and is averaged over 25 largest momentum and
    charge symmetric subspaces. Full curves show the Wigner surmises for the
    Poissonian, circular orthogonal (COE) and circular unitary (CUE) ensemble~\cite{Haake}.}
    \label{fig:level_spacing}
\end{figure}

The considered systems \eqref{eq:block_gate} and  \eqref{eq:block_gate_3site}
exhibit an $r$-site translational symmetry and conserve their respective charges
$\hat Q$, defined in \eqref{eq:charge_stag} and \eqref{eq:charge_3site}. An eigenbasis of both
of these symmetry operators can simply be constructed by taking sums of
translations of the computational basis. Additionally, both systems exhibit a
so-called space-time symmetry. Namely, in both cases the propagator can be
written as a power of its ``root'', $\hat{\mathbb{U}} = \hat{\mathbb{V}}^r \hat
T^{-r}$, where $\hat{\mathbb{V}} = \hat{\mathbb U}_o \hat T$ in the $r = 2$
spin-1 case and $\hat{\mathbb{V}} = \hat{\mathbb{U}}_0 \hat T$ in the $r = 3$
spin-$\tfrac12$ case. As shown in
Ref.~\onlinecite{duhClassificationSamegateQuantum2024}, in such systems, the root
$\hat{\mathbb{V}}$ is expected to show RMT level-spacings, whereas the whole
propagator $\mathbb{U}$ exhibits the level-spacing of the appropriate RMT
ensemble ``wound around'' the unit circle $r$ times. Resolving all the
symmetries in the considered systems, therefore, amounts to studying the
levels-spacing distribution of the root $\mathbb{V}$ in irreducible symmetry
subspace of translational symmetry and charge $\hat Q$. Numerical calculations
of the said level-spacings are shown in Fig.~\ref{fig:level_spacing}, where we
see that spin-1 case follows the predictions of the circular unitary ensemble,
signaling chaotic behavior without time-reversal (anti-unitary) symmetries, and
the spin-$\tfrac12$ case follows the predictions of the circular orthogonal
ensemble, signalling chaotic behavior with an additional time-reversal symmetry.

\SMsection{``Low-charge'' excitations for three-site qubit gate}
In this section we will derive the ``low-charge'' physics of the 3 site qubit gate \eqref{eq:block_gate_3site} presented in the main text. The conserved charge is given by $\hat{Q}=\sum_n \left(2\hat{S}^z_{3n}+\hat{S}^z_{3n+1}+\hat{S}^z_{3n+2}\right)$. As $\mu \to \infty$, we reach the state with the highest charge $\ket{\Omega}=\ket{\cdots 00000\cdots}$,  such that $Q_{max}=2N$. We define the charge deficit $\hat{D}:=2N-\hat{Q}$, such that subspace of $D=0$ is given by state $\ket{\Omega}$. Let us denote the basis states with exactly one $1$ at site $j$ and $0$ elsewhere by $\ket{j}$. Then, the subspace with lowest charge excitation $D=1$ is then spanned by the states
\begin{align}
    \ket{3j+1},\ket{3j+2}; \hspace{1cm} j\in\mathbb{Z}.
\end{align}
A flip at site $3j$ belongs to the $D=2$ sector.

The three-site translation symmetry is resolved by the Fourier states
\begin{align}
    \ket{k,z} &:= \sum_j e^{3ikj} \ket{3j+z}; \hspace{1cm} z\in\qty{1,2}, k\in (-\pi/3,\pi/3].
\end{align}
We can write $\hat{\mathbb{U}} = \hat{T}^{3}(\hat{T}^\dagger\hat{\mathbb{U}}_0)^3$. We have $\hat{T} \ket{k, z} = \ket{k, z}$ by construction. Therefore, if we can diagonalise $\hat{T}^\dagger\hat{\mathbb{U}}_0$ within the subspace $\{\ket{k, z}\}_{z=1,2}$, we diagonalise $\hat{\mathbb{U}}$. Now consider the action of $\hat{T}^\dagger\hat{\mathbb{U}}_0$ on $\ket{k, z}$:
\begin{align}
    \hat{T}^\dagger\hat{\mathbb{U}}_0\ket{k,1} &= \sum_j e^{3ikj} \hat{T}^\dagger\hat{\mathbb{U}}_0\ket{3j+1} = \sum_j e^{3ikj} V_{001\leftarrow 001}\ket{3j+1} + \sum_j e^{3ikj} V_{100\leftarrow 001}\ket{3j-1}\\
    &=  V_{001\leftarrow 001} \ket{k,1} + V_{100\leftarrow 001} e^{3ik} \ket{k,2}\\
    \hat{T}^\dagger \hat{\mathbb{U}}_0\ket{k,2} &= \sum_j e^{3ikj} \hat{T}^\dagger\hat{\mathbb{U}}_0\ket{3j+2} = \sum_j e^{3ikj} V_{001\leftarrow 100}\ket{3j+1} + \sum_j e^{3ikj} V_{100\leftarrow 100}\ket{3j-1}\\
    &=  V_{001\leftarrow 100} \ket{k,1} + V_{100\leftarrow 100} e^{3ik} \ket{k,2}.
\end{align}
Thus, for each $k$ we can find the eigenstates by diagonalizing the $2\times 2$ unitary matrix
\begin{align}
    \hat{A}_k^\top &= \begin{pmatrix}
        V_{001\leftarrow 001} & V_{100\leftarrow 001} e^{3ik}\\
        V_{001\leftarrow 100} & V_{100\leftarrow 100} e^{3ik}
    \end{pmatrix}  = \hat{V}_{\{001, 100\}} \begin{pmatrix}
            1 & 0\\
            0 & e^{3ik}
        \end{pmatrix}.
\end{align}
The two eigenvalues are phases, which we write as $e^{-i\varepsilon_k^\pm +ik}$, s.t.\ the corresponding eigenvalues of $\hat{\mathbb{U}}$ are $e^{-3i\varepsilon_k^\pm}$. For the specific model used in the main text with gate \eqref{eq:block_gate_3site}, the $\varepsilon_k^\pm$ are plotted in Fig. \ref{fig:sm_let}.
\begin{figure}
    \centering
    \includegraphics[width=0.5\linewidth]{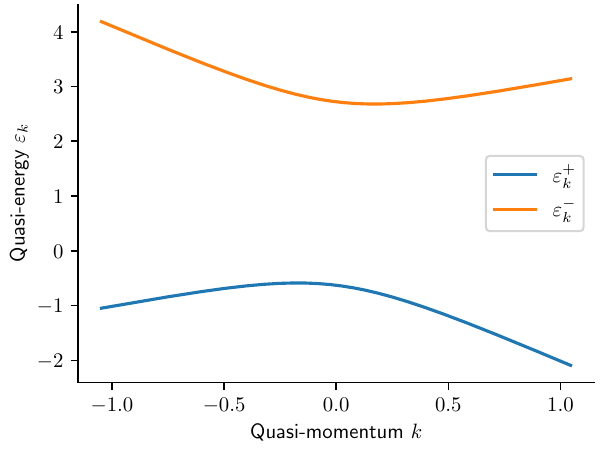}
    \caption{The two branches of the quasienergy $\varepsilon_k^\pm$ as function of quasimomentum $k$ for the circuit \eqref{eq:block_gate_3site}.}
    \label{fig:sm_let}
\end{figure}
At this point, we can view the evolution in this single particle sector to be that of a two-band system with dispersion relations $\varepsilon_k^\pm$. In particular, wavepackets will move with the group velocity $v^\pm(k)=\dv{k}\varepsilon_k^\pm$. To see this, let us denote the eigenvectors of $\hat{A}_k$ by $\vec{w}^\pm(k) = (w^\pm_1(k),w^\pm_2(k))^\top$ and evaluate the correlation function for large $\mu$ as follows.
\begin{align}
    \expval{\hat{q}_{t,j}\hat{q}_{0}}\upd{c} &= \expval{(\hat{q}_{t,j}-2)(\hat{q}_{0}-2)}\upd{c}\\
    &= \frac{\tr \qty[e^{\mu\hat{Q}}(\hat{q}_{t,j}-2)(\hat{q}_{0}-2)]}{\tr\qty[e^{\mu \hat{Q}}]}-\qty(\frac{\tr \qty[e^{\mu\hat{Q}}(\hat{q}_{0}-2)]}{\tr \qty[e^{\mu \hat{Q}}]})^2\\
    &= \frac{\tr \qty[e^{-\mu\hat{D}}(\hat{q}_{t,j}-2)(\hat{q}_{0}-2)]}{\tr \qty[e^{-\mu \hat{D}}]}-\qty(\frac{\tr \qty[e^{-\mu\hat{D}}(\hat{q}_{0}-2)]}{\tr \qty[e^{-\mu \hat{D}}]})^2,
\end{align}
where $\hat{D}:=2N-\hat{Q}$ as above. Now, we can expand each local density matrix factor in $e^{-\mu \hat{D}} = \prod_x (\hat\rho^{x \, \mathrm{mod} \, 3}_\mu)_x$ and collect the powers of $e^{-\mu}\ll 1$ to retrieve
\begin{align}
    e^{-\mu \hat{D}} &= \ket{\Omega}\bra{\Omega} + e^{-\mu}\sum_j \qty(\ket{3j+1}\bra{3j+1}+\ket{3j+2}\bra{3j+2}) + \order{e^{-2\mu}}.
\end{align}
Since $(\hat{q}_0-2)\ket{\Omega} = 0$, we can further simplify
\begin{align}
    (\hat{q}_0-2)e^{-\mu \hat{D}} &= -e^{-\mu} \qty(\ket{1}\bra{1}+\ket{2}\bra{2}) + \order{e^{-2\mu}}.
\end{align}
Therefore, to leading order in $e^{-\mu}$, we obtain
\begin{align}
    \expval{\hat{q}_{t,j}\hat{q}_{0}}\upd{c} &=-e^{-\mu}\tr \qty[(\hat{q}_{t,j}-2)\qty(\ket{1}\bra{1} + \ket{2}\bra{2}) + \order{e^{-2\mu}}]\\
    &= -e^{-\mu}\sum_{z=1}^2\bra{z} \hat{\mathbb{U}}^{-t}(\hat{q}_{j}-2)\hat{\mathbb{U}}^t\ket{z} + \order{e^{-2\mu}}\\
    &= e^{-\mu}\sum_{z,z'=1}^2\bra{z} \hat{\mathbb{U}}^{-t}\ket{3j+z'}\bra{3j+z'}\hat{\mathbb{U}}^t\ket{z} + \order{e^{-2\mu}}\\
    &= e^{-\mu}\sum_{z,z'=1}^2\abs{\bra{3j+z'}\hat{\mathbb{U}}^t\ket{z}}^2 + \order{e^{-2\mu}}\\
    &= e^{-\mu}\sum_{z,z'=1}^2\abs{\int_{-\pi/3}^{\pi/3}\frac{\dd{k}}{2\pi/3} \bra{3j+z'}\hat{\mathbb{U}}^t\ket{k,z}}^2 + \order{e^{-2\mu}}\\
    &= e^{-\mu}\sum_{z,z'=1}^2\abs{\sum_{\sigma=+,-}\int_{-\pi/3}^{\pi/3}\frac{\dd{k}}{2\pi/3} e^{3ikj-3i\varepsilon_k^\sigma t} w^\sigma_{z'}(k)w^{\sigma*}_{z}(k)}^2 + \order{e^{-2\mu}}.\label{eq:effective_theory}
\end{align}

For large $t$ and $j$ we can approximate these integrals using a stationary phase approximation, which generally predicts a decay as $1/\qty(\partial_k^2\varepsilon^\pm_k t)$. From this, it is clear that the dominant contribution comes from $k$ where $\partial_k^2
\varepsilon^\pm_k =0$. Let us assume that there is only one \footnote{In case of multiple such $k$ one simply has to take a sum over all the different contributions.} of such $k$ which we denote by $k^\pm$. We can then write
\begin{align}
    \int_{-\pi/3}^{\pi/3}\frac{\dd{k}}{2\pi/3} &e^{3ikj-3i\varepsilon_k^\sigma t} w^\sigma_{z'}(k)w^{\sigma*}_{z}(k) \approx e^{3ik^\sigma -3i\varepsilon_{k^\sigma}^\sigma t}\int\frac{\dd{k}}{2\pi/3} e^{3i(k-k^\sigma)j-3iv^\sigma(k^\sigma) (k-k^\sigma) t - \tfrac{1}{2}i\partial^3_k\varepsilon_{k^\sigma}^\sigma (k-k^\sigma)^3 t} w^\sigma_{z'}(k)w^{\sigma*}_{z}(k)\\
    &\approx e^{3ik^\sigma -3i\varepsilon_{k^\sigma}^\sigma t}w^\sigma_{z'}(k^\sigma)w^{\sigma*}_{z}(k^\sigma)\int\frac{\dd{k}}{2\pi/3} e^{3ikj-3iv^\sigma(k^\sigma) k t - \tfrac{1}{2}i\partial^3_k\varepsilon_{k^\sigma}^\sigma k^3 t}\\
    &\approx e^{3ik^\sigma -3i\varepsilon_{k^\sigma}^\sigma t}w^\sigma_{z'}(k^\sigma)w^{\sigma*}_{z}(k^\sigma)\int_0^\infty\frac{\dd{k}}{\pi/3} \cos(3(j-v^\sigma(k^\sigma)t)k - \tfrac{1}{2}\partial^3_k\varepsilon_{k^\sigma}^\sigma k^3 t)\\
    &= e^{3ik^\sigma -3i\varepsilon_{k^\sigma}^\sigma t}w^\sigma_{z'}(k^\sigma)w^{\sigma*}_{z}(k^\sigma)\sqrt[3]{\frac{18}{\abs{\partial_k^3\varepsilon^\sigma_{k^\sigma}} t}}\mathrm{Ai}\qty(-\sgn(\partial_k^3\varepsilon^\sigma_{k^\sigma})\sqrt[3]{\frac{18}{\abs{\partial_k^3\varepsilon^\sigma_{k^\sigma}} t}}(j-v^\sigma(k^\sigma)t)).
\end{align}
Inserting this into the above expression, we find
\begin{align}
    \expval{\hat{q}_{t,j}\hat{q}_{0}}\upd{c} &\approx e^{-\mu}\sum_{z,z'=1}^2\sum_{\sigma=+,-}\abs{w^\sigma_{z'}(k^\sigma)}^2\abs{w^{\sigma}_{z}(k^\sigma)}^2\qty(\frac{18}{\abs{\partial_k^3\varepsilon^\sigma_{k^\sigma}} t})^{\tfrac{2}{3}}\mathrm{Ai}^2\qty(-\sgn(\partial_k^3\varepsilon^\sigma_{k^\sigma})\sqrt[3]{\frac{18}{\abs{\partial_k^3\varepsilon^\sigma_{k^\sigma}} t}}(j-v^\sigma(k^\sigma)t))\\
    &= e^{-\mu}\sum_{\sigma=+,-}\qty(\frac{18}{\abs{\partial_k^3\varepsilon^\sigma_{k^\sigma}} t})^{\tfrac{2}{3}}\mathrm{Ai}^2\qty(-\sgn(\partial_k^3\varepsilon^\sigma_{k^\sigma})\sqrt[3]{\frac{18}{\abs{\partial_k^3\varepsilon^\sigma_{k^\sigma}} t}}(j-v^\sigma(k^\sigma)t)),
\end{align}
as stated in the End Matter of the main text.

\SMsection{Details of numerical simulations}

To demonstrate our predictions, we numerically probed the dynamical
autocorrelation functions of the charge density in the thermodynamic limit
\begin{align}
    C(x, t) := \ev{\hat \rho_\mu \hat q(x, t) \hat q(0, 0)} = \ev{\hat \rho_\mu \hat q(0, t) \hat q(-x, 0)}.
\end{align}
In order to do this in a finite-size simulation, we must perform the numerics on
a large enough number of sites $L$, such that the superdiffusive causal cone
spreading as $x \sim (\lambda_Bt)^{2/3}$ has not yet reached the boundary. In
order to reach $L \sim 4800$ in our simulations, we represent
$\hat q(0, 0)$ as a matrix-product operator (MPO) and propagate it with
time-evolving block decimation
(TEBD)~\cite{vidalEfficientClassicalSimulation2003,
vidalEfficientSimulationOnedimensional2004}, keeping only a finite bond
dimension $\chi$ and truncating the highly entangled components. Schematically
denoting the TEBD-propagated charge density as $\mathcal U_\mathrm{TEBD}^t[\hat
q(0, 0)]$, we effectively approximate $\hat q(0, t) \approx \mathcal
U_\mathrm{TEBD}^t[\hat q(0, 0)]$ and thus 
\begin{equation}
    C(x, t) \approx \ev{\hat \rho_\mu \mathcal U_\mathrm{TEBD}^t[\hat q(0, 0)] \hat q(-x, 0)}. \label{eq:tebd_inf}
\end{equation}
We find that such an approximation works well for infinite temperature $\hat
\rho_{\mu = 0} \sim \hat{\mathbb{I}}$, e.g., in Fig.~\ref{fig:Corr_profile}. As
a simple error estimate, we note that the integral of the autocorrelation
function $\sum_x C(x, t)$, which is conserved for exact unitary evolution,
deviates from its initial value $\sum_x C(x, 0)$ by $10^{-3}$ over the entire
course of the simulation.

The prescription \eqref{eq:tebd_inf} fails for finite temperature (nonzero
chemical potential) $\mu > 0$, e.g., we find that $\sum_x C(x, t)$ deviates
from $\sum_x C(x, 0)$ by a few percent already at $t = 20$
for the cases considered in Fig.~\ref{fig:qubit_profiles}. In order to obtain accurate results for large $\mu$, we
employ two modifications. Firstly, the truncation procedure in TEBD is geared
towards minimizing the error in the Hilbert-Schmidt norm, i.e., one can bound
$\norm{\hat q(0, t) - \mathcal U_\mathrm{TEBD}^t[\hat q(0, 0)]}$, where
$\norm{A} := \sqrt{\ev{A^\dagger A}}$, with the discarded Schmidt coefficients. While
this is suitable for infinite temperature, since one contracts $\ev{\hat q(0, t)
\hat q(-x, 0)}$ in the Hilbert-Schmidt inner product, the inner product at $\mu
> 0$ is weighted with $\rho_\mu$. To remedy this, we symmetrize the inner
product and numerically simulate $\sqrt{\hat \rho_\mu} \hat q(0, t) \approx \mathcal
U_\mathrm{TEBD}^t\left[\sqrt{\hat \rho_\mu} \hat q(0, 0)\right]$. Heuristically, in this case,
one obtains a good approximation of $\sqrt{\hat
\rho_\mu} \hat q(0, t)$ in the Hilbert-Schmidt norm, which one then contracts with $\sqrt{\hat \rho_\mu} \hat
q(-x, 0)$ in the corresponding (Hilbert-Schmidt) inner product. We note that a 
conceptually similar approach was already used before, see, e.g., Ref.~\onlinecite{srivatsaProdbingHydro2026}.

Secondly, obtaining a good approximation for the entire operator $\hat q(0, t)$
(we omit $\sqrt{\hat \rho_\mu}$ for now to lighten the notation) in the
Hilbert-Schmidt norm is neither necessary nor sufficient for accurately
predicting autocorrelation functions of local observables. Namely, since we later
contract $\hat q(0, t)$ with the local $\hat q(-x, 0)$, we must actually
accurately predict the local components of $\hat q(0,
t)$~\cite{rakovszkyDAOE2022, keyserlingkBackflow2022,
duh2026quantummanybodyoperatorcascade}. To obtain that, we employ an idea
reminiscent of the domain-wall quench approach found to work well in numerically
obtaining local correlation functions~\cite{ljubotina2017spin, Ljubotina2019}. That is, instead of
propagating $\hat q(0, 0)$ with TEBD, we propagate $\hat{\mathbb{I}} + \epsilon
q(0, 0)$ for some small $\epsilon$~\footnote{This approach was suggested to us by Marko
Ljubotina}. That this approach indeed prefers local operator strings in the truncation step
of TEBD at infinite temperature will be clarified in an upcoming manuscript~\cite{ljubotina_toapp}.

Putting both ideas together, we numerically simulate
\begin{equation}
    \epsilon C(x, t) \approx \ev{\mathcal U_\mathrm{TEBD}^t\left[\sqrt{\hat \rho_\mu} \left(\hat{\mathbb{I}} + \epsilon \hat q(0, 0)\right)\right]\sqrt{\hat \rho_\mu} \hat q(-x, 0)} - \ev{\hat \rho_\mu \hat q(0, 0)}.
\end{equation}
Employing such an approach, we find that $\abs{\sum_x C(x, t) - \sum_x C(x, 0)} < 10^{-4}$
for all the cases considered in the main text
(Figs.~\ref{fig:qubit_profiles} and \ref{fig:FWHM}) using $\epsilon = 10^{-4}$.

On a technical note, we remark also that simulating $\mathcal
U_\mathrm{TEBD}^t\left[\sqrt{\hat \rho_\mu} \, \right]$ with open boundary
conditions will incur a large error at one of the boundaries already after 1
time step, since our charge density ``drifts'' after each time step
(cf.~Eq.~\eqref{eq:q_drift}). We remedy this by resetting the boundaries back to
$\sim \sqrt{\hat \rho_\mu}$ by hand after every time step.

\end{document}